\documentclass[11pt,a4paper]{article}

\usepackage{jheppub}
\usepackage{physics}
\usepackage{braket} % To use \bra{text} and \ket{text}.
\usepackage{relsize} % To use \mathsmaller.
\usepackage{etoolbox} % To use \ifblank etc.
\usepackage{color} % To use \color{color}.
\usepackage[bottom]{footmisc} % To push footnote to the very bottom of page.

\renewcommand{\imaginary}{i}
\renewcommand{\exp}[1]{e^{#1}}
\renewcommand{\Re}[1]{\text{Re}{\left(#1\right)}}

\newcommand{\inMode}[1]{\psi_{#1}^{\mathsmaller{\text{in}}}}
\newcommand{\inAnOp}[2]{#1_{#2}^{\mathsmaller{\text{in}}}}
\newcommand{\inCrOp}[2]{#1_{#2}^{\mathsmaller{\text{in}}\dagger}}

\newcommand{\inVacBra}{\bra{0_{\text{in}}}}
\newcommand{\inVacKet}{\ket{0_{\text{in}}}}

\newcommand{\psir}{\psi_{+}} % Coefficient of \Up.
\newcommand{\psil}{\psi_{-}} % Coefficient of \Um.
\newcommand{\Up}{U_{+}}	% Eigenstate of \alpha with eigenvalue +1.
\newcommand{\Um}{U_{-}}	% Eigenstate of \alpha with eigenvalue -1.
\newcommand{\Upm}{U_{\pm}} % Eigenstate of \alpha with eigenvalue \pm.
\newcommand{\Ump}{U_{\mp}} % Eigenstate of \alpha with eigenvalue \mp.

\newcommand{\sgn}[1]{\mathop\text{sgn}(#1)}	% Signum function.
\newcommand{\BigO}[1]{O\!\left(#1\right)}
\newcommand{\BigOinf}[1]{O^{\infty}\!\left(#1\right)}
\DeclareDocumentCommand{\F}{ O{} O{} }{F_{\mathsmaller{\text{#1}}}^{#2}(\omega)} % The response function.
\DeclareDocumentCommand{\Fconst}{ O{} O{} }{F_{\mathsmaller{\text{#1}}}^{#2}} % The response function constant term.

\newcommand{\BbbR}{\mathbb{R}}

\begin{document}

% Title
\title{The cost of building a wall for a fermion}

% Authors
\author[a,b]{Wan Mohamad Husni Wan Mokhtar}
\author[b]{and Jorma Louko}

\affiliation[a]{School of Physics, Universiti Sains Malaysia,
11800 USM, Penang, Malaysia} 

\affiliation[b]{School of Mathematical Sciences, University of Nottingham, Nottingham NG7 2RD, UK}

% Email
\emailAdd{wanhusni@usm.my, jorma.louko@nottingham.ac.uk}

\abstract{We analyse the energy cost of building or demolishing a wall for a massless Dirac field in (1+1)-dimensional Minkowski spacetime and the response of an Unruh-DeWitt particle detector to the generated radiation. For any smoothly-evolving wall, both the field's energy density and the detector's response are finite. In the limit of rapid wall creation or demolition, the energy density displays a delta function squared divergence. By contrast, the response of an Unruh-DeWitt detector, evaluated within first-order perturbation theory, diverges only logarithmically in the duration of the wall evolution. The results add to the evidence that a localised matter system may not be as sensitive to the rapid wall creation as the local expectation values of field observables. This disparity has potential interest for quantum information preservation scenarios.}

\maketitle

\section{Introduction}

In quantum field theory, time dependence in the spacetime metric or 
in the field's boundary conditions generically results 
in particle creation accompanied by an energy flow \cite{Mukhanov:2007zz,Birrell:1982ix}. 
Well-known examples include particle creation 
in an expanding universe \cite{Parker:1968mv,Parker:1969au,Parker:1971pt,Parker:2012at}, 
particle creation by black hole formation 
\cite{Hawking:1974sw} and particle creation by one or more moving 
boundaries \cite{Moore:1970,Davies:1976hi,Candelas:1977zza}. 
The last example, known as dynamical Casimir effect, 
has been observed experimentally, using waveguides and Josephson 
junctions to simulate mechanical motion~\cite{Wilson:2011,Lahteenmaki:2013mda}. 

In this paper we consider the response of a quantum field 
to the smooth or sudden creation or demolition of a boundary. 
One motivation is that this process provides a model for the dynamical quantum 
correlation breaking that is the main ingredient 
in the black hole firewall argument \cite{Almheiri:2012rt} 
(for debate with references, see~\cite{Marolf:2017jkr,Unruh:2017uaw}). 
The process may also capture aspects of the bulk-boundary interaction that occurs when 
the boundary is given dynamical 
degrees of freedom~\cite{G.:2015yxa,Barbero:2017kvi,Dappiaggi:2018pju}. 
Mathematically, the creation or demolition of a boundary can be implemented as 
time evolution of the field under an explicitly time-dependent Hamiltonian, 
and this evolution has been shown to admit a description within 
algebraic quantum field theory~\cite{Benini:2017dfw}.  

For a massless scalar field in $(1+1)$-dimensional Minkowski spacetime, the creation of a static two-sided Dirichlet wall was analysed in~\cite{Brown:2015yma}, following the earlier perturbative analysis in~\cite{Obadia:2001hj}. In the rapid wall creation limit, the energy emitted into the field on the light cone of the creation event was found to diverge, but the response of an inertial Unruh-DeWitt detector \cite{Unruh:1976db,DeWitt:1980hx} was nevertheless found to remain finite. 
In this process, a rapid creation of the wall hence affects a localised matter 
system coupled to the field less strongly than what would be expected from the divergence in the local energy density. 
A~similar conclusion was obtained in \cite{Louko:2014aba,Martin-Martinez:2015dja} 
for a quantum state in which spatial correlations across Rindler horizons are broken by hand. 
Related observations in terms of the particle content after the wall 
has formed were made in~\cite{Brown:2014qna}. In the context where the mirror-field interaction is modelled using a non-perturbative circuit model, the entanglement between the energy bursts emanating from the mirror was analysed in~\cite{Foo:2020rsy}.
An instantaneous wall demolition was analysed in~\cite{Harada:2016kkq}, 
showing that also this process is energetically singular, 
although less strongly so than a wall creation. 

All the above analyses address a massless scalar field in $1+1$ dimensions. 
While this allows significant technical simplifications, 
it also leaves open to what extent the conclusions may generalise to our 
$(3+1)$-dimensional home spacetime, 
and to fields that are experimentally known to exist. 
One point of concern is that a massless scalar field in $1+1$ 
dimensions requires an infrared cutoff, 
and this cutoff still appears in several of the bottom line results, 
including those in~\cite{Brown:2015yma}. 
Another point of concern is that quantum field theories tend to 
become more singular as the spacetime dimension increases. 
For a massless scalar field in $3+1$ dimensions, 
it was found in \cite{Zhou:2016hsh,Carrington:2018ikq} 
that the creation of a spherical wall that only affects 
the spherically symmetric sector of the field is broadly 
similar to the $(1+1)$-dimensional wall creation of~\cite{Brown:2015yma}, 
but the creation of a pointlike source is indeed significantly more singular, 
both energetically and for the response of an inertial detector. 

The purpose of this paper is to analyse how a Dirac field reacts  
to the creation and demolition of a wall. 
Is there again a disparity between energetic considerations 
and the effects felt by a localised matter system? 

We work in $(1+1)$-dimensional Minkowski spacetime and consider a massless two-component Dirac field, taking advantage of the technical simplifications that this affords~\cite{Takagi:1986kn,Louko:2016ptn,Hummer:2015xaa}. 
The evolving wall is mechanically static at the spatial origin. 
At each moment of time, the wall is described by a choice in the $\mathrm{U}(2)$ family of self-adjointness boundary conditions, and the evolution of the wall is implemented by allowing the boundary condition to depend on time. Some, but not all, of these boundary conditions have an interpretation in terms of an idealised non-local separable potential~\cite{Calkin:1988zz}.

For comparability with the scalar field analyses in \cite{Brown:2015yma,Brown:2014qna,Harada:2016kkq}, 
we consider the smooth creation and demolition of an impermeable wall that imposes on each side the MIT bag boundary condition~\cite{Chodos:1974je}, and the short time limit of this process 
where the creation and demolition become instantaneous. 

First, we show that the renormalised stress-energy tensor of the Dirac field is finite everywhere 
away from a smoothly created or demolished wall. 
In the rapid creation or demolition limit, the total energy injected into the field diverges, and while the details of the divergence depend on how the limit is taken, we exhibit a class of limiting prescriptions in which the divergence is proportional to $\delta^{-1}$ where $\delta$ is the duration of the process. The energy of the Dirac field is hence better behaved than the energy of the scalar field, in that no infrared regulator is needed, and when the evolution takes place over a strictly finite interval of time, there is no divergence of the kind that was observed for the scalar field in the post-publication note in the eprint v4 of \cite{Brown:2015yma}. The divergence of the total energy in the rapid evolution limit is nevertheless qualitatively similar to that of the scalar field. 

Second, we show that a pointlike detector, 
coupled linearly to the scalar density of the Dirac field and treated within first-order perturbation theory, 
has a finite response to a smoothly created or demolished wall. 
In the rapid creation or demolition limit, the response however diverges, logarithmically in the duration of the process. This is a striking contrast with the scalar field case~\cite{Brown:2015yma}, where the (infrared-regulated) response remained finite in the rapid evolution limit. 

In summary, these results show that the rapid creation or demolition of a two-sided MIT bag wall is singular both energetically and for the response of a localised matter system, but the singularity in the latter is weaker. 
This adds to the evidence that a localised matter system may not be as sensitive to the rapid wall creation as the local expectation values of field observables. 
More broadly, it adds to the evidence that quantum information preservation scenarios, 
such as the black hole firewall argument~\cite{Marolf:2017jkr,Unruh:2017uaw}, 
need to be investigated by carefully specifying the means by which the quantum field is probed. 

The organisation of the paper is as follows. We begin in Section \ref{Section Classical Field} by introducing general time-dependent $\mathrm{U}(2)$ boundary conditions at the wall, and we single out certain one-parameter subgroups, enumerated by an integer, for the wall creation and demolition analysis. We also discuss the relation of these boundary conditions to an idealised delta-potential, and we present the mode solutions to the field equation. In Section \ref{Section Quantum Field}, we quantise the field and analyse the renormalised energy density. The response of an inertial Unruh-DeWitt particle detector is analysed in Section \ref{Section Particle Detector}, first for a general time-dependent wall, and then for wall creation and demolition over a finite interval of time and in the rapid creation limit. The main part of the paper ends with a discussion in Section \ref{Section Discussion}. 
Technical details are deferred to two appendices.

We denote complex conjugation by an asterisk. $\BigO{x}$ denotes a quantity such that $\BigO{x}\!/x$ remains bounded as $x \to 0$, 
and $\BigOinf{x}$ denotes a quantity that goes to zero faster than any positive power of $x$ as $x \to 0$. We employ units in which $\hbar = c = 1$.

\section{Classical Field} \label{Section Classical Field}
	
	\subsection{General Boundary Condition} \label{Section General Boundary Condition}
	
We work in $1+1$ Minkowski spacetime, in the standard global coordinates $(t,z)$ 
in which the metric reads $\dd{s}^{2} = \dd{t}^{2} - \dd{z}^{2}$. 
The Dirac equation for our two-component spinor field $\psi$ 
takes the form 
	\begin{equation} \label{Dirac Equation}
		\imaginary \partial_{t} \psi = H \psi,
	\end{equation}
where $H = - \imaginary \alpha \partial_{z} + m \beta$ is the Dirac operator, $m$ is the mass of the field and $\alpha, \beta$ are $2 \times 2$ hermitian anti-commuting matrices that square to unity.
The Hilbert space is $L^{2}(\mathbb{R}) \oplus L^{2}(\mathbb{R})$, with the inner product 
\begin{align} \label{Inner Product}
		(\psi_{1},\psi_{2}) 
			&= \int_{-\infty}^{\infty} [\psi_{1}(z)]^\dagger\psi_{2}(z) \dd{z}.
	\end{align}

We implement the split of $\psi$ into its two components by \cite{Friis:2011yd,Friis:2013eva}
\begin{align} \label{Basis Expansion}
\psi = \psir\Up + \psil\Um, 
\end{align}
where the spinor basis 
$\{\Up,\Um\}$ is orthonormal, 
in the sense that $\Up^\dagger\Um = 0$ and $\Up^\dagger\Up = \Um^\dagger\Um = 1$, and satisfies 
\begin{align}
\alpha\Upm = \pm\Upm, \qquad
\beta\Upm = \Ump.
\end{align}	
In the representation in which 
$\alpha = 
\big(\begin{smallmatrix}
  1 & 0\\
  0 & -1
\end{smallmatrix}\big)$
and 
$\beta = 
\big(\begin{smallmatrix}
  0 & 1\\
  1 & 0
\end{smallmatrix}\big)$, 
we may choose 
$U_+ = 
\big(\begin{smallmatrix}
  1\\
  0
\end{smallmatrix}\big)$
and 
$U_- = 
\big(\begin{smallmatrix}
  0\\
  1
\end{smallmatrix}\big)$, 
in which case 
$\psi = 
\big(\begin{smallmatrix}
  \psi_+\\
  \psi_-
\end{smallmatrix}\big)$; the choice of the representation will however not affect what follows. 

We specialise from now on to a massless field, $m = 0$. 
Substituting \eqref{Basis Expansion} in the Dirac equation \eqref{Dirac Equation} 
then shows that $\psir$ is a right-mover and $\psil$ is a left-mover: in terms of the global Minkowski null coordinates $u := t-z$ and $v := t+z$, we have $\psir = \psir(u)$ and $\psil = \psil(v)$. 

To introduce a wall that is mechanically motionless but has time-dependent internal structure, 
we replace the Dirac equation \eqref{Dirac Equation} by 
	\begin{equation} \label{Dirac Equation Extensions}
		\imaginary \partial_{t} \psi = H_{U(t)} \psi,
	\end{equation}
where the time-dependent Hamiltonian $H_{U(t)}$
coincides with the usual massless Dirac Hamiltonian on each of the half-lines $z>0$ and $z<0$, 
but glues these half-lines together at $z=0$ by a time-dependent boundary 
condition that is general enough to allow partial reflection and transmission, 
while ensuring that $H_{U(t)}$ is self-adjoint at each~$t$. 
As outlined in Appendix \ref{Appendix Self-adjoint Extensions}, 
the most general such boundary condition reads 
	\begin{align} \label{General Wall Boundary Conditions 1}
		\begin{pmatrix}
			\psir(t,0_{+}) \\
			\psil(t,0_{-})
		\end{pmatrix}
		&= U(t)
		\begin{pmatrix}
			\psir(t,0_{-}) \\
			\psil(t,0_{+})
		\end{pmatrix} , 
	\end{align}
where $\psi_\pm$ are as in~\eqref{Basis Expansion}, 
$\psir(t,0_{\pm}) := \lim_{z \to 0_{\pm}} \psir(t,z)$, $\psil(t,0_{\pm}) := \lim_{z \to 0_{\pm}} \psil(t,z)$, 
and $U(t) \in \mathrm{U}(2)$ is a $t$-dependent unitary $2\times2$ matrix. 

We assume that $U(t)$ is smooth in~$t$.  
We also assume that $U(t)$ is time-dependent only over a finite interval of time, $t\in[t_0, t_0 + \delta]$, where the positive parameter $\delta$ is the interval's length: 
$U(t)$ is a constant matrix for $t\le t_0$ and a (possibly different) constant matrix for $t\ge t_0 + \delta$. 

$\mathrm{U}(2)$ is four-dimensional. 
A matrix $U \in \mathrm{U}(2)$ may be parametrised by four real-valued parameters 
$(\theta, \varphi_1, \varphi_2, \varphi)$ as 
	\begin{align} \label{General Wall Boundary Conditions 2}
		U &=
		\exp{\imaginary\varphi}
		\begin{pmatrix}
			\exp{\imaginary\varphi_{1}} \cos\theta
				& \imaginary\exp{\imaginary\varphi_{2}} \sin\theta \\
			\imaginary\exp{-\imaginary\varphi_{2}} \sin\theta
				& \exp{-\imaginary\varphi_{1}} \cos\theta
		\end{pmatrix}, 
	\end{align}
where $0\le \theta \le \pi/2$, 
$\varphi_1$, $\varphi_2$ and $\varphi$ are each understood periodic with period $2\pi$, and the overall identification 
$(\theta, \varphi_1, \varphi_2, \varphi) \sim (\theta, \varphi_1 + \pi, \varphi_2 + \pi, \varphi + \pi)$ is understood. The parametrisation has a coordinate singularity at $\theta=0$, where $\varphi_2$ is ambiguous, and at $\theta=\pi/2$, where $\varphi_1$ is ambiguous; these coordinate singularities are similar to the coordinate singularity of the planar polar coordinates at the origin, 
with $\sin\theta$ and $\cos\theta$ playing the role of the radial coordinate. 
$U(t)$~can hence be specified by specifying the four parameters  
$(\theta, \varphi_1, \varphi_2, \varphi)$ as functions of~$t$.

	\subsection{Wall Creation and Demolition} \label{Section Wall Appearance and Disappearance}

We now turn to the choice of $U(t)$ before and after the evolution. 

One choice of interest is $U_{\text{NW}} := I$. By \eqref{General Wall Boundary Conditions 1}, this 
makes $\psi$ continuous across $z=0$, which means that there is no wall: 
the Hamiltonian is the usual massless Dirac field Hamiltonian. 
The subscript ``NW'' stands for ``no wall''. 

A second choice of interest is 
	\begin{align} \label{Static MIT wall boundary condition 2}
		U_{\text{MIT}} := \begin{pmatrix} 0 & -i \\ -i & 0 \end{pmatrix}.
	\end{align}	
By \eqref{General Wall Boundary Conditions 1}, this 
reads 
	\begin{align} \label{Static MIT wall boundary condition in spinor basis}
		\psi_{\pm}(t,0_{\pm}) = - i \psi_{\mp}(t,0_{\pm}), 
	\end{align}
which fully decouples the field at $z>0$ from the field at $z<0$: 
there is an impermeable wall at $z=0$. 
The condition \eqref{Static MIT wall boundary condition in spinor basis} can be written as 
\begin{align} 
\label{Static MIT wall boundary condition 1}
(1 \mp i \beta\alpha) \psi(t,0_{\pm}) = 0, 
\end{align}
or as 
\begin{align}
i n_{\mu} \gamma^{\mu} \psi\big|_{\text{boundary}}
= \psi\big|_{\text{boundary}},
\label{General MIT bag boundary condition} 
\end{align}
where $\gamma^\mu = \{\beta,\beta\alpha\}$ are the gamma-matrices and $n_{\mu}$ is, on each side of the wall, the unit normal that points inward from the wall: $n_{\mu} = (0,1)$ for $z\to 0_+$ and $n_{\mu} = (0,-1)$ for $z\to 0_-$. 
This is the MIT bag boundary condition, 
found in \cite{Chodos:1974je} as the limiting case of a wall separating fields with different masses when the mass on the other side of the wall is taken to infinity. 
Among all totally reflective boundary conditions for fermions, the MIT bag condition 
can be seen as the analogue of the Dirichlet condition for scalar fields~\cite{Belchev:2010}. 

The creation of an MIT bag wall is hence described by $U(t)$ 
that evolves from $U_{\text{NW}}$ to $U_{\text{MIT}}$; 
conversely, the demolition of an MIT bag wall is described by $U(t)$ 
that evolves from $U_{\text{MIT}}$ to~$U_{\text{NW}}$. 
We shall initially consider general paths of evolution in~$\mathrm{U}(2)$, 
but later specialise to paths along one-parameter subgroups in~$\mathrm{U}(2)$, 
of the form 
\begin{align} \label{Discrete Family of Boundary Conditions}
		U_{n}(t)
		&=
		\exp{2 \imaginary n\theta(t)}
		\begin{pmatrix}
			\cos\theta(t) & (-1)^{n+1} \imaginary\sin\theta(t) \\
			(-1)^{n+1} \imaginary\sin\theta(t) & \cos\theta(t)
		\end{pmatrix},
	\end{align}
where the discrete parameter $n \in \mathbb{Z}$ specifies the subgroup. 
In terms of the parametrisation in \eqref{General Wall Boundary Conditions 2}, 
the paths \eqref{Discrete Family of Boundary Conditions} are recovered with  
$\varphi(t) = 2n\theta(t)$, $\varphi_{1}(t) = 0$ and $\varphi_{2}(t) = (n+1)\pi$. 
$U_{\text{NW}}$ is obtained at $\theta = 0$, and $U_{\text{MIT}}$ is obtained at $\theta = \pi/2$. 
The creation of an MIT wall over time interval $\delta$ is hence described by 
	\begin{align} \label{Condition of Theta for Wall Appearance}
		\theta(t) = \theta_{1}(t) :=
			\begin{cases}
				0 & \text{for } t \leq t_{0}, \\
				h_{1}(t) & \text{for } t_{0} < t < t_{0} + \delta, \\
				\frac{\pi}{2} & \text{for } t \geq t_{0} + \delta,
			\end{cases}
	\end{align}
while the demolition of an MIT wall over time interval $\delta$ is described by 
	\begin{align} \label{Condition of Theta for Wall Disappearance}
		\theta(t) = \theta_{2}(t) :=
			\begin{cases}
				\frac{\pi}{2} & \text{for } t \leq t_{0}, \\
				h_{2}(t) & \text{for } t_{0} < t < t_{0} + \delta, \\
				0 & \text{for } t \geq t_{0} + \delta,
			\end{cases}
	\end{align}
where $h_{1}(t)$ and $h_{2}(t)$ are smooth interpolating functions. 
A sudden creation or demolition of the wall is obtained in the limit 
$\delta \to 0$. 

We note here in passing, and demonstrate in more detail in Appendix~\ref{Appendix Delta Potential}, 
that it is possible to interpret some of the boundary conditions 
\eqref{General Wall Boundary Conditions 2} 
in terms of a Dirac equation that includes a potential 
proportional to $\delta(z)$ with a time-dependent coefficient, so that Dirac's equation reads 
	\begin{align} \label{Dirac Equation Delta Potential General}
		\imaginary\partial_t\psi
			= -\imaginary\alpha\partial_{z}\psi + \bigl(S(t)\beta+V(t)\bigr)\delta(z)\psi , 
	\end{align}
where the presciption
	\begin{align} 
		\lim_{\epsilon \to 0_{+}} \int_{-\epsilon}^{+\epsilon} \delta(z)\psi(z) \dd{z}
			:= \frac{1}{2}\bigl(\psi(0_{-}) + \psi(0_{+})\bigr) \label{Delta Potential Common Definition}
	\end{align}	
is understood. 
In particular, the wall creation and demolition via the evolution 
\eqref{Discrete Family of Boundary Conditions}--\eqref{Condition of Theta for Wall Disappearance}
is obtained with
\begin{align} \label{S(t) and V(t)}
		S(t) = \frac{2(-1)^{n}\sin\theta(t)}{\cos\theta(t) + \cos{2n\theta(t)}}, \qquad
		V(t) = -\frac{2\sin{2n\theta(t)}}{\cos\theta(t) + \cos{2n\theta(t)}},
	\end{align}
where $0 \leq \theta(t) \leq \pi/2$. 
Note, however, that if $n\ne0$ in~\eqref{S(t) and V(t)}, 
both $S(t)$ and $V(t)$
diverge somewhere along the evolution between 
$U_{\text{NW}}$ and~$U_{\text{MIT}}$. 
This suggests that providing a physical model for the internal dynamics of 
the wall may be easier for $n=0$ than for $n\ne0$.

\subsection{Mode Solutions}

We now turn to the mode solutions for the field. 

Recall that as the field is massless, we have $\psir = \psir(u)$ and $\psil = \psil(v)$. 
We hence seek mode functions by the ansatz 
	\begin{align}
		\inMode{\text{r},k}(u,v) &=
			\begin{cases}
				\frac{1}{\sqrt{2\pi}} \exp{-\imaginary ku} \Up + E_{\text{r},k}(v) \Um 
					& \text{for } z < 0, \\
				F_{\text{r},k}(u) \Up 
					& \text{for } z > 0,
			\end{cases} \label{In-mode Solution R Ansatz} \\
		\inMode{\text{l},k}(u,v) &=
			\begin{cases}
				E_{\text{l},k}(v) \Um 
					& \text{for } z < 0, \\
				F_{\text{l},k}(u) \Up + \frac{1}{\sqrt{2\pi}} \exp{-\imaginary kv} \Um 
					& \text{for } z > 0,
			\end{cases} \label{In-mode Solution L Ansatz}
	\end{align}
where $k\in\mathbb{R}\setminus\{0\}$ and the functions 
$E_{\text{r},k}(v)$, 
$F_{\text{r},k}(u)$, 
$E_{\text{l},k}(v)$
and 
$F_{\text{l},k}(u)$ 
are to be determined from the junction conditions at $z=0$. 
$\inMode{\text{r},k}(u,v)$ describes a pure right-moving plane wave, 
proportional to $\exp{-\imaginary ku} \Up$, coming in from $z<0$, 
and then partially reflected and partially transmitted at the wall; 
similarly, 
$\inMode{\text{l},k}(u,v)$ 
describes a pure left-moving plane wave, 
proportional to $\exp{-\imaginary kv} \Um$, coming in from $z>0$, 
and then partially reflected and partially transmitted at the wall. 

Substituting the ansatz \eqref{In-mode Solution R Ansatz}--\eqref{In-mode Solution L Ansatz} 
in \eqref{General Wall Boundary Conditions 1}, 
with the general boundary condition~\eqref{General Wall Boundary Conditions 2}, 
we find 
	\begin{align}
		\inMode{\text{r},k}(u,v) &=
			\begin{cases}
				\frac{1}{\sqrt{2\pi}} \exp{-\imaginary ku} \Up 
				+ \frac{\imaginary}{\sqrt{2\pi}} \exp{-\imaginary kv + \imaginary\varphi(v) - \imaginary\varphi_{2}(v)} \sin\theta(v) \Um
					& \text{for } z < 0, \\
				\frac{1}{\sqrt{2\pi}} \exp{-\imaginary ku + \imaginary\varphi(u) + \imaginary\varphi_{1}(u)} \cos\theta(u) \Up
					& \text{for } z > 0,
			\end{cases} \label{In-mode Solution R} \\
		\inMode{\text{l},k}(u,v) &=
			\begin{cases}
				\frac{1}{\sqrt{2\pi}} \exp{-\imaginary kv + \imaginary\varphi(v) - \imaginary\varphi_{1}(v)} \cos\theta(v) \Um
					& \text{for } z < 0, \\
				\frac{\imaginary}{\sqrt{2\pi}} \exp{-\imaginary ku + \imaginary\varphi(u) + \imaginary\varphi_{2}(u)} \sin\theta(u) \Up
				+ \frac{1}{\sqrt{2\pi}} \exp{-\imaginary kv} \Um
					& \text{for } z > 0.
			\end{cases} \label{In-mode Solution L}
	\end{align}
It can be verified that the mode solutions are \eqref{In-mode Solution R}--\eqref{In-mode Solution L} are (Dirac) orthonormal in the inner product \eqref{Inner Product Punctured}, in the sense of $( \inMode{\sigma,k}, \inMode{\sigma',k'}) = \delta_{\sigma\sigma'} \delta(k-k')$, where $\sigma,\sigma' 
\in \{\text{r},\text{l}\}$. 
As the functions 
$\bigl(\theta(t), \varphi_{1}(t), \varphi_{2}(t), \varphi(t)\bigr)$ 
are constants for 
$t \leq t_{0}$, before the wall starts to evolve, 
$\inMode{\text{r},k}$ and $\inMode{\text{l},k}$ 
reduce to stationary mode solutions in the spacetime region $t-|z| \le t_0$, 
which is causally disconnected from the evolution of the wall, and which we may call the in-region. 
In particular, if there is no wall at $t\le t_0$, then
$\inMode{\text{r},k}(u,v) \propto \exp{-\imaginary ku} \Up$ and $\inMode{\text{l},k}(u,v) \propto \exp{-\imaginary kv} \Um$ everywhere in the in-region, 
making $\inMode{\text{r},k}(u,v)$ a pure right-moving plane wave 
and $\inMode{\text{l},k}(u,v)$ is a pure left-moving plane wave, with $k$ the wave number.

	\section{Quantum Field} \label{Section Quantum Field}
	
	\subsection{Quantisation} \label{Section Quantisation}
	
Given the mode solutions \eqref{In-mode Solution R}--\eqref{In-mode Solution L}, 
we expand the quantised Dirac field as 
	\begin{align} \label{General Solution Time-Dependent Wall}
		\psi(u,v) 
			&= \int_{0}^{\infty} \dd{q}
				{\left( \inAnOp{a}{\text{r},q}\inMode{\text{r},q} + \inAnOp{a}{\text{l},q}\inMode{\text{l},q} 
				+ \inCrOp{b}{\text{r},q}\inMode{\text{r},-q} + \inCrOp{b}{\text{l},q}\inMode{\text{l},-q} \right)}, 
	\end{align}
where the operators 
$\inAnOp{a}{\text{r},q}$, $\inAnOp{a}{\text{l},q}$, $\inAnOp{b}{\text{r},q}$ and $\inAnOp{b}{\text{l},q}$ have by construction a positive momentum index~$q$, 
and their nonvanishing anticommutators are 
\begin{align} \label{Canonical Anticommutation Relation}
		\left\{\inAnOp{a}{\sigma,p},\inCrOp{a}{\sigma',q}\right\}
			= \left\{\inAnOp{b}{\sigma,p},\inCrOp{b}{\sigma',q}\right\}
			= \delta_{\sigma\sigma'}\delta(p-q). 
\end{align}
The operators $\inAnOp{a}{\text{r},q}$, $\inAnOp{a}{\text{l},q}$, $\inAnOp{b}{\text{r},q}$ and $\inAnOp{b}{\text{l},q}$ 
are then the annihilation operators corresponding to the in-modes \eqref{In-mode Solution R}--\eqref{In-mode Solution L}, and their Hermitian conjugates are the corresponding creation operators. 
The in-vacuum $\inVacKet$ is the normalised state that is annihilated by these annihilation operators, satisfying 
	\begin{align} \label{In-vacuum definition}
	\inAnOp{a}{\text{r},k} \inVacKet = \inAnOp{a}{\text{l},k} \inVacKet
		= \inAnOp{b}{\text{r},k} \inVacKet = \inAnOp{b}{\text{l},k} \inVacKet = 0 . 
%			\qquad \text{for } k > 0.
\end{align}
In the in-region, $t-|z| \le t_0$, $\inVacKet$ 
has the physical interpretation as the no-particle state in the presence of a static wall.

\subsection{Field Propagators}

We now consider the field propagators, 
which will be needed for analysing both the stress-energy and the response of an Unruh-DeWitt detector. 

The positive and negative frequency propagators in the state $\inVacKet$ are defined as \cite{Takagi:1986kn}
	\begin{align}
		S_{ab}^{+}(u,v;u',v')
			&:= \inVacBra \psi_{a}(u,v) \overline{\psi}_{b}(u',v') \inVacKet,
				\label{Evolving wall positive frequency propagator} \\
		S_{ab}^{-}(u,v;u',v')
			&:= \inVacBra \overline{\psi}_{b}(u',v') \psi_{a}(u,v) \inVacKet,
				\label{Evolving wall negative frequency propagator}
	\end{align}
where $\overline{\psi} := \psi^\dagger\beta$ is the Dirac conjugate of $\psi$, and the 
subscripts indicate the components of the two-component spinors. 
Substituting the mode expansion \eqref{General Solution Time-Dependent Wall}
and the mode solutions \eqref{In-mode Solution R}--\eqref{In-mode Solution L} in 
\eqref{Evolving wall positive frequency propagator}--\eqref{Evolving wall negative frequency propagator}, we find for $z,z'>0$ the expressions 
\begin{align}
		&S_{ab}^{+}(u,v;u',v') \nonumber \\
			&\qquad =
				- \frac{i}{2\pi} \frac{1}{v - v' - i \epsilon} U_{-,a} U_{-,c}^{\dagger} \beta_{cb}
				+ \frac{1}{2\pi} \frac{ e^{i\varphi(u) + i\varphi_{2}(u)} \sin\theta(u) }
					{u - v' - i \epsilon} U_{+,a} U_{-,c}^{\dagger} \beta_{cb}
						\nonumber \\
			&\qquad \qquad
				- \frac{1}{2\pi} \frac{ e^{-i\varphi(u') - i\varphi_{2}(u')} \sin\theta(u') }
					{v - u' - i \epsilon} U_{-,a} U_{+,c}^{\dagger} \beta_{cb}
						\nonumber \\
			&\qquad	\qquad
				- \frac{i}{2\pi} \frac{ e^{i[\varphi(u)-\varphi(u')] + i[\varphi_{2}(u)-\varphi_{2}(u')]}
					\sin\theta(u) \sin\theta(u') } {u - u' - i \epsilon} U_{+,a} U_{+,c}^{\dagger} \beta_{cb}
						\nonumber \\
			&\qquad	\qquad
				- \frac{i}{2\pi} \frac{ e^{i[\varphi(u)-\varphi(u')] + i[\varphi_{1}(u)-\varphi_{1}(u')]}
					\cos\theta(u) \cos\theta(u') } {u - u' - i \epsilon} U_{+,a} U_{+,c}^{\dagger} \beta_{cb},
						\label{Evolving wall Splus distribution right-hand side}
	\end{align}% Manual layout
	\begin{align}
		&S_{ab}^{-}(u,v;u',v') \nonumber \\
			&\qquad =
				\frac{i}{2\pi} \frac{1}{v - v' + i \epsilon} U_{-,a} U_{-,c}^{\dagger} \beta_{cb}
				- \frac{1}{2\pi} \frac{ e^{i\varphi(u) + i\varphi_{2}(u)} \sin\theta(u) }
					{u - v' + i \epsilon} U_{+,a} U_{-,c}^{\dagger} \beta_{cb}
						\nonumber \\
			&\qquad \qquad
				+ \frac{1}{2\pi} \frac{ e^{-i\varphi(u') - i\varphi_{2}(u')} \sin\theta(u') }
					{v - u' + i \epsilon} U_{-,a} U_{+,c}^{\dagger} \beta_{cb}
						\nonumber \\
			&\qquad	\qquad
				+ \frac{i}{2\pi} \frac{ e^{i[\varphi(u)-\varphi(u')] + i[\varphi_{2}(u)-\varphi_{2}(u')]}
					\sin\theta(u) \sin\theta(u') } {u - u' + i \epsilon} U_{+,a} U_{+,c}^{\dagger} \beta_{cb}
						\nonumber \\
			&\qquad	\qquad
				+ \frac{i}{2\pi} \frac{ e^{i[\varphi(u)-\varphi(u')] + i[\varphi_{1}(u)-\varphi_{1}(u')]}
					\cos\theta(u) \cos\theta(u') } {u - u' + i \epsilon} U_{+,a} U_{+,c}^{\dagger} \beta_{cb},
						\label{Evolving wall Sminus distribution right-hand side}
	\end{align}
where $i\epsilon$ indicates the distributional limit $\epsilon \to 0_{+}$,  
arising from the distributional interpretation of the integral \cite{Birrell:1982ix,Mukhanov:2007zz}
\begin{align} \label{Distributional interpretation}
		\int_{0}^{\infty} \dd{p} e^{i p w}
			\to \lim_{\epsilon \to 0_{+}} \int_{0}^{\infty} \dd{p} e^{i p w - p \epsilon}
%			= \lim_{\epsilon \to 0_{+}} \frac{1}{\epsilon - i w}
			= \lim_{\epsilon \to 0_{+}} \frac{i}{w + i \epsilon} . 
%			= i P \left(\frac{1}{w}\right) + \pi \delta(w) .
\end{align}
The distribution in \eqref{Distributional interpretation} can be written as 
$i P \! \left(\frac{1}{w}\right) + \pi \delta(w)$, 
where $P$ stands for the Cauchy principal value, but for us the $i \epsilon$ 
representation will be more convenient. 
Similar expressions can be obtained when the two points are in other parts of the spacetime.

	\subsection{Stress-Energy Tensor}

We can now turn to the energy that is transmitted into the quantum field by the evolving wall. 

Recall that the stress-energy tensor $T_{\mu\nu}$ for a classical Dirac field $\psi$ is given by \cite{Birrell:1982ix}
	\begin{align} \label{Stress Energy Tensor}
		T_{\mu\nu}
			= \frac{\imaginary}{2} \left[ \, \overline{\psi} \gamma_{(\mu} \partial_{\nu)} \psi
				- \left( \partial_{(\mu} \overline{\psi} \, \right) \gamma_{\nu)} \psi \right],
	\end{align}
where $\gamma_{\mu} := \eta_{\mu\nu}\gamma^{\nu}$, 
and $A_{(\mu\nu)} := \frac12 (A_{\mu\nu}+A_{\nu\mu})$ is the symmetric part of $A_{\mu\nu}$. 
For the quantised field, the expectation value of the stress-energy tensor in the state $\inVacKet$ requires a renormalisation, 
which we do by point-splitting and subtracting the corresponding expression in Minkowski vacuum~\cite{Birrell:1982ix}. 
Focusing on the energy density, 
$\expval{T_{tt}}_{\text{in}} := \inVacBra T_{tt} \inVacKet$, 
we find that the point-split version of 
$\expval{T_{tt}}_{\text{in}}$ is \cite{WanMokhtar:2018lwi}
\begin{align}
	\expval{T_{tt}(u,v; u',v')}_{\text{in}}^{\text{split}}
			= \frac{i}{2} \Big[ \partial_{u} F(u,v;u',v') + \partial_{v} F(u,v;u',v') \Big],
				\label{Ttt formula}
\end{align}
where
\begin{align}
		F(u,v;u',v') &:= \Tr{S^{-}(u,v;u',v')\beta} - \Tr{S^{-}(u',v';u,v)\beta}, 
			\label{F for stress-energy tensor}
	\end{align}
and the trace is over the spinor indices.
The corresponding expression for Minkowski vacuum, $\expval{T_{tt}(u,v; u',v')}_{\text{Mink}}^{\text{split}}$, is obtained by setting $\theta = \varphi = \varphi_1 = \varphi_2 =0$. 
The renormalised energy density is then obtained as the coincidence limit of the difference, 
\begin{align}
\expval{T_{tt}}_{\text{in}}^{\text{ren}}
= 
 \lim_{(u',v') \to (u,v)}
 \left( \lim_{\epsilon \to 0_+}\left( \expval{T_{tt}(u,v; u',v')}_{\text{in}}^{\text{split}} - \expval{T_{tt}(u,v; u',v')}_{\text{Mink}}^{\text{split}} \right) \right) , 
\end{align}
where we have explicitly indicated that
the $\epsilon\to0_+$ 
limit in our $i\epsilon$ notation is taken under the coincidence limit 
since the subtraction cancels the distributional singularities. 

For $z>0$, \eqref{Evolving wall Sminus distribution right-hand side} now gives 
	\begin{align} \label{General Time-Dependent Vacuum Energy Density Expectation Value Right}
		\expval{T_{tt}}_{\text{in}, \, z>0}^{\text{ren}}
			&=
				\frac{1}{4\pi} \left[ \bigl(\varphi'(u)\bigr)^{2}
				+ \bigl(\varphi'_{1}(u)\bigr)^{2} \cos^{2} \! \theta(u)
				+ \bigl(\varphi'_{2}(u)\bigr)^{2} \sin^{2} \! \theta(u)
				+ \bigl(\theta'(u)\bigr)^{2} \right]
					\nonumber \\
			&\qquad
				+ \frac{1}{2\pi} \varphi'(u)
					\left[ \varphi'_{1}(u) \cos^{2} \! \theta(u)
					+ \varphi'_{2}(u) \sin^{2} \! \theta(u) \right] , 
	\end{align}
and a similar calculation for $z<0$ gives 
	\begin{align} \label{General Time-Dependent Vacuum Energy Density Expectation Value Left}
		\expval{T_{tt}}_{\text{in}, \, z<0}^{\text{ren}}
			&=
				\frac{1}{4\pi} \left[ \bigl(\varphi'(v)\bigr)^{2}
				+ \bigl(\varphi'_{1}(v)\bigr)^{2} \cos^{2} \! \theta(v)
				+ \bigl(\varphi'_{2}(v)\bigr)^{2} \sin^{2} \! \theta(v)
				+ \bigl(\theta'(v)\bigr)^{2} \right]
					\nonumber \\
			&\qquad
				- \frac{1}{2\pi} \varphi'(v)
					\left[ \varphi'_{1}(v) \cos^{2} \! \theta(v)
					+ \varphi'_{2}(v) \sin^{2} \! \theta(v) \right] , 
	\end{align}
where the primes on $\theta$, $\varphi$, $\varphi_1$ and $\varphi_2$ 
denote derivative with respect to the argument. 

Three observations are in order. 

First, 
$\expval{T_{tt}}_{\text{in}}^{\text{ren}}$ at a spacetime point depends on the wall's evolution only at the null separated point on the wall. This had to happen because the field is massless. 

Second, 
since the wall evolves smoothly by assumption, $\expval{T_{tt}}_{\text{in}}^{\text{ren}}$ 
is well defined and smooth both for $z<0$ and for $z>0$. It does not need an infrared regulator, like the corresponding massless scalar field in $1+1$ dimensions~\cite{Brown:2015yma}. It does not diverge on the light cone of points where the wall parameters approach special values, 
like for the corresponding scalar field in $1+1$ dimensions (see the post-publication note in the eprint v4 of \cite{Brown:2015yma}) and in $3+1$ dimensions~\cite{Zhou:2016hsh,Carrington:2018ikq}. While the parametrisation of $U(t)$ by \eqref{General Wall Boundary Conditions 2} contains a coordinate singularity at $\theta=0$, where $\varphi_2$ becomes ambiguous, and at $\theta=\pi/2$, where $\varphi_1$ becomes ambiguous, these coordinate singularities are similar to those of planar polar coordinates at the origin, and the combinations appearing in 
\eqref{General Time-Dependent Vacuum Energy Density Expectation Value Right}--\eqref{General Time-Dependent Vacuum Energy Density Expectation Value Left} 
are smooth whenever $U(t)$ is smooth in the standard differentiable structure of~$\mathrm{U}(2)$. 

Third, 
point-splitting is crucial to correctly subtract the divergent Minkowski contribution to the stress-energy tensor. 
Proceeding without point-splitting would give for $\expval{T_{tt}}_{\text{in}}$ expressions that are divergent, 
for $z>0$ given by 
	\begin{align} \label{General Time-Dependent Vacuum Energy Density Expectation Value Right Alt}
		\expval{T_{tt}}_{\text{in}}
			= \int_{0}^{\infty} \frac{\dd{p}}{2\pi}
				\left( -2p - \varphi'(u) - \varphi'_{1}(u)\cos^2 \! \theta(u)
					- \varphi'_{2}(u)\sin^2 \! \theta(u) \right) , 
	\end{align}
and for $z<0$ by 
	\begin{align} \label{General Time-Dependent Vacuum Energy Density Expectation Value Left Alt}
		\expval{T_{tt}}_{\text{in}}
			= \int_{0}^{\infty} \frac{\dd{p}}{2\pi}
				\left( -2p - \varphi'(v) + \varphi'_{1}(v)\cos^2 \! \theta(v)
					+ \varphi'_{2}(v)\sin^2 \! \theta(v) \right) , 
	\end{align}
and subtracting under the integral the `Minkowski contribution,' 
obtained with $\theta = \varphi = \varphi_1 = \varphi_2 =0$, does not give agreement with 
\eqref{General Time-Dependent Vacuum Energy Density Expectation Value Right}--\eqref{General Time-Dependent Vacuum Energy Density Expectation Value Left}; indeed, it gives expressions that are still ultraviolet divergent for generic wall evolution, while vanishing when the only evolving wall parameter is~$\theta$, both of which properties are in disagreement with 
\eqref{General Time-Dependent Vacuum Energy Density Expectation Value Right}--\eqref{General Time-Dependent Vacuum Energy Density Expectation Value Left}. 
Point-splitting is demonstrably crucial. 
	
	\subsection{Wall Creation and Demolition} \label{Section Wall Appearance and Disappearance Stress Energy Tensor}

We now consider the wall creation and demolition through the families 
\eqref{Discrete Family of Boundary Conditions}
with \eqref{Condition of Theta for Wall Appearance}--\eqref{Condition of Theta for Wall Disappearance}. 

Substituting $\varphi(t) = 2n\theta(t)$, $\varphi_{1}(t) = 0$ and $\varphi_{2}(t) = (n+1)\pi$ in \eqref{General Time-Dependent Vacuum Energy Density Expectation Value Left}--\eqref{General Time-Dependent Vacuum Energy Density Expectation Value Right}, 
we find
\begin{align} \label{Renormalised Vacuum Energy Density for Wall Appearance and Disappearance}
		\expval{T_{tt}}_{\text{in}}^{\text{ren}} = \frac{1+4n^{2}}{4\pi} \times
		\begin{cases}
			 \bigl[\theta'(v) \bigr]^{2}
					&\text{for } z < 0, \\
			\bigl[\theta'(u) \bigr]^{2}
					&\text{for } z > 0.
		\end{cases}
	\end{align}
As noted above, $\expval{T_{tt}}_{\text{in}}^{\text{ren}}$ is well defined everywhere. 
We use \eqref{Renormalised Vacuum Energy Density for Wall Appearance and Disappearance} to examine what happens in the limit of instantaneous wall creation or demolition, $\delta \to 0$. 

First, under mild assumptions about the $\delta$-dependence of the interpolating functions $h_1$ and $h_2$ in 
\eqref{Condition of Theta for Wall Appearance}--\eqref{Condition of Theta for Wall Disappearance}, 
the $\delta\to0$ limit gives 
$\theta'(t) \to (\pi/2) \delta(t-t_{0})$ for wall creation 
and 
$\theta'(t) \to - (\pi/2) \delta(t-t_{0})$ for wall demolition. 
As the square of the Dirac delta does not exist as a distribution, 
this shows that $\expval{T_{tt}}_{\text{in}}^{\text{ren}}$ 
is not defined on the light cone of the event at which the wall instantaneously appears or disappears. 

Second, consider the total energy that the wall creation or demolition ejects into the field, given by 
	\begin{align} \label{GEW expectation value of total energy}
		\expval{E} = \int_{-\infty}^{\infty} \expval{T_{tt}}_{\text{in}}^{\text{ren}} \dd{z},
	\end{align}
evaluated on any constant $t$ hypersurface at $t > t_{0} + \delta$. 
From \eqref{Renormalised Vacuum Energy Density for Wall Appearance and Disappearance} and \eqref{Condition of Theta for Wall Appearance}--\eqref{Condition of Theta for Wall Disappearance}, we find 
	\begin{align} \label{WCD expectation value of total energy}
		\expval{E}
			= \frac{1+4n^{2}}{2\pi} \int_{0}^{\delta} \bigl[h'_{j}(t_{0}+x)\bigr]^{2} \dd{x} , 
	\end{align}
where $j=1$ is for wall creation and $j=2$ is for wall demolition. 
To establish a lower bound for $\expval{E}$, 
we note that the Cauchy-Schwarz inequality applied to the functions 
$h'_{j}(t_{0}+x)$ and $1$ gives 
\begin{align} 
\label{Wall creation or demolition Cauchy-Schwarz inequality}
\abs{ \int_{0}^{\delta} h'_{j}(t_{0}+x) \cdot 1 \dd{x} }^{2}
			\leq \int_{0}^{\delta} \abs{h'_{j}(t_{0}+x)}^{2} \dd{x} \cdot \int_{0}^{\delta} \abs{1}^{2} \dd{x} , 
	\end{align}
where the left-hand side evaluates to $\pi^2/4$ because 
\begin{align}
\int_{0}^{\delta} h'_{j}(t_{0}+x) \dd{x}
			= h_{j}(t_{0}+\delta) - h_{j}(t_{0}) 
			= 
\begin{cases}
\pi/2 & \text{for} \ j=1 , 
\\
- \pi/2 & \text{for} \ j=2 , 
\end{cases}\end{align}
using \eqref{Condition of Theta for Wall Appearance}--\eqref{Condition of Theta for Wall Disappearance}. 
We hence have 
\begin{align}
\expval{E}
\ge 
\frac{\pi \bigl(1+4n^{2}\bigr)}{8\delta}  , 
\end{align}
which shows that the total energy emitted into the field diverges as $\delta\to0$, at least 
proportionally to $\delta^{-1}$. 

As a special case, suppose that the $\delta$-dependence of $\theta_j$ 
is an overall scaling of the independent variable. 
We then have 
\begin{align}
\theta_j(t) = g_j \bigl( (t-t_0)/\delta \bigr),  
\end{align}
where $g_j$ are $\delta$-independent smooth monotonic functions, 
satisfying 
$g_1(y) =0$ for $y\le0$ and $g_1(y) =\pi/2$ for $y\ge1$, 
and 
$g_2(y) = \pi/2$ for $y\le0$ and $g_2(y) = 0$ for $y\ge1$. 
From \eqref{WCD expectation value of total energy} we then have 
\begin{align} \label{WCD expectation value of total energy with delta-scaling}
\expval{E}
= \frac{1+4n^{2}}{2\pi \delta} \int_{0}^{1} \bigl[g'_{j}(y)\bigr]^{2} \dd{y} . 
\end{align}
$\expval{E}$ is hence a multiple of $1/\delta$, 
by a coefficient that depends on the detailed form of~$g_j$. 
By the above Cauchy-Schwarz argument, 
the coefficient is bounded below by $\pi \bigl(1+4n^{2}\bigr)/8$.

\section{Unruh-DeWitt Detector} \label{Section Particle Detector}

In this section we consider the response of an Unruh-DeWitt detector 
in the spacetime of the evolving wall.

\subsection{The Response Function\label{subsec:responsefunction}}

Our Unruh-DeWitt detector is a spatially pointlike quantum system, 
moving on the timelike worldline $\bigl(t(\tau),z(\tau)\bigr)$, 
where the parameter $\tau$ is the proper time. 
The internal dynamics is that of a two-level system, 
where the energies are defined with respect to~$\tau$. 
Without loss of generality, we may choose one energy eigenvalue to be~$0$, 
and we denote the other by $\omega \in \BbbR \setminus \{0\}$. 

We take the detector to couple to the quantum field with the interaction Hamiltonian \cite{Takagi:1986kn,Louko:2016ptn,Hummer:2015xaa} 
	\begin{align} \label{Scalar density interaction Hamiltonian}
		H_{\text{int}}(\tau) = c \mu(\tau) \chi(\tau) \overline{\psi}\bigl(u(\tau),v(\tau)\bigr) \psi\bigl(u(\tau),v(\tau)\bigr),
	\end{align}
where $c$ is a real-valued coupling constant, $\mu(\tau)$ is the detector's monopole moment operator and $\chi(\tau)$ is a real-valued switching function that specifies how the interaction is switched on and off. We assume $\chi$ to be smooth with compact support.

Consider an interaction scenario in which, before the interaction begins, 
the field is prepared in the state~$\inVacKet$, 
described in Section~\ref{Section Quantum Field}, 
and the detector is prepared in the state with energy~$0$. 
In first-order perturbation theory, the probability for the detector to be found in the state with energy $\omega$ after the interaction has ceased, regardless the final state of the field, 
is a multiple of the response function~$\F$, given by \cite{Louko:2016ptn}
\begin{align} \label{Detector Response Function}
		\F = \int_{-\infty}^{\infty} \dd{\tau} \int_{-\infty}^{\infty} \dd{\tau'}
					\exp{-\imaginary\omega(\tau-\tau')} \chi(\tau) \chi(\tau') W^{(2,\bar{2})}(\tau,\tau') , 
\end{align}
where $W^{(2,\bar{2})}(\tau,\tau') := W^{(2,\bar{2})}\bigl(u(\tau),v(\tau);u(\tau'),v(\tau')\bigr)$ is the pull-back on the detector's worldline of the field's two-point correlation function, 
	\begin{align}
		W^{(2,\bar{2})}(u,v;u',v')
			&:= \inVacBra \overline{\psi}(u,v) \psi(u,v) \overline{\psi}(u',v') \psi(u',v') \inVacKet . 
\label{eq:W-field-correlation}
	\end{align}
The constant of proportionality involves only $c^2$ and a matrix element of $\mu(0)$, encoding the detector's internal structure,
but all the dependence on $\inVacKet$, $\chi$, $\omega$ and the detector's trajectory 
is contained in~$\F$. 
With minor abuse of terminology, common in the literature, 
we may refer to $\F$ as the transition probability. 

The two-point correlation function $W^{(2,\bar{2})}$ \eqref{eq:W-field-correlation} can be written in terms of the field propagators $S^{\pm}$ 
\eqref{Evolving wall positive frequency propagator}--\eqref{Evolving wall negative frequency propagator}
as \cite{Louko:2016ptn}
	\begin{align} \label{Vacuum two-point function in terms of propagators}
		W^{(2,\bar{2})}(u,v;u',v')
			&= \Tr{S^{+}(u,v;u',v')S^{-}(u',v';u,v)} \nonumber \\
			&\qquad + \Tr{S^{-}(u,v;u,v)}\Tr{S^{-}(u',v';u',v')}.
	\end{align}
While $S^{\pm}$ are well-defined distributions, 
$W^{(2,\bar{2})}$ involves products of distributions, which are not a priori well defined; 
further, the second term in \eqref{Vacuum two-point function in terms of propagators}
involves $S^{-}$ evaluated at the coincidence limit singularity, 
which is certainly not well defined. 
However, when the matrix trace in \eqref{Vacuum two-point function in terms of propagators} is taken, 
what remains turns out to be well defined; in particular, 
from \eqref{Evolving wall Sminus distribution right-hand side} 
we find \cite{WanMokhtar:2018lwi}
\begin{align} 
\Tr{S^{-}(u,v;u,v)}
= 
- \frac{\sin\theta(u)}{2\pi}
\left(
\frac{e^{i\varphi(u) + i \varphi_2(u)}}{u - v + i \epsilon}
+ 
\frac{e^{-i\varphi(u) - i \varphi_2(u)}}{u - v - i \epsilon}
\right) , 
\label{eq:Sminus-coincident-trace}
\end{align}
which is a well-defined distribution. 
We shall interpret $W^{(2,\bar{2})}$ this way, 
using the $\epsilon$-representation for~$S^{\pm}$, 
and taking the limit $\epsilon\to0$ at the end. 

We note in passing that an alternative to 
$H_{\text{int}}$ 
\eqref{Scalar density interaction Hamiltonian}
would be the normal-ordered interaction Hamiltonian
\begin{align} \label{Normally ordered scalar density interaction Hamiltonian}
H_{\text{int}}^{\text{no}}(\tau) = c \mu(\tau) \chi(\tau) :\overline{\psi}(u(\tau),v(\tau)) \psi(u(\tau),v(\tau)): 
\ , 
\end{align}
where $:\overline{\psi}\psi:$ denotes normal ordering~\cite{Hummer:2015xaa}. 
For~$H_{\text{int}}^{\text{no}}$, 
$W^{(2,\bar{2})}$ differs from 
\eqref{Vacuum two-point function in terms of propagators}
only in that the second term is absent. 
From \eqref{eq:Sminus-coincident-trace} we see that 
$H_{\text{int}}$ and $H_{\text{int}}^{\text{no}}$ give the same results when there is no wall, but they do not do so for generic wall evolutions. 
We shall work primarily with~$H_{\text{int}}$, 
but we shall make occasional comments on what would differ for~$H_{\text{int}}^{\text{no}}$. 

From now on we take the detector's trajectory to be parallel to the wall, 
at $z= d>0$. 
We parametrise the trajectory as 
	\begin{align} \label{Detector Trajectory}
		\bigl(t(\tau),z(\tau)\bigr) = (\tau + d, d). 
	\end{align}
In this parametrisation, 
the information about the wall's evolution at the event $(t_{\text{event}}, 0)$ 
reaches the detector at detector proper time $\tau = t_{\text{event}}$.

\subsection{No Wall} \label{Section Absence of Wall}
	
We start by recalling the detector's response in the absence of a wall, $\F[NW]$. 
By Theorem 1 of~\cite{Louko:2016ptn}, $\F[NW]$ is twice the response of an inertial 
Unruh-DeWitt detector that is coupled linearly to a scalar field in $3+1$ Minkowski spacetime. 
From (3.8) in \cite{Satz:2006kb} we have 
\begin{align} % \label{FNW}
		\F[NW]
			&= -\frac{\omega}{2\pi} \int_{-\infty}^{\infty} \dd{u} [\chi(u)]^2
				+ \frac{1}{\pi^2} \int_{0}^{\infty}  \frac{\dd{s}}{s^2}
						\int_{-\infty}^{\infty} \dd{u} \chi(u)[\chi(u)-\chi(u-s)]
\notag
\\[1ex]
& \hspace{3ex}
+ \frac{1}{\pi^2} \int_{-\infty}^{\infty} \dd{u} \chi(u)
\int_0^{\infty}
\dd{s} \chi(u-s) \, \frac{1 - \cos\omega{s}}{s^2} . 
\end{align}
To obtain a more transparent form, we interchange the integrals in the last term, 
write $\chi(u) \chi(u-s) = \chi(u) [\chi(u-s) - \chi(u)] + [\chi(u)]^2$,
and use the identity 
$\int_{0}^{\infty} (\sin(\omega{s})/s) \dd{s} = (\pi/2) \sgn{\omega}$. 
Combining terms gives
\begin{align} \label{FNW}
		\F[NW]
			&= -\frac{\omega\Theta(-\omega)}{\pi} \int_{-\infty}^{\infty} \dd{u} [\chi(u)]^2
				+ \frac{1}{\pi^2} \int_{0}^{\infty} \dd{s} \frac{\cos\omega{s}}{s^2}
						\int_{-\infty}^{\infty} \dd{u} \chi(u)[\chi(u)-\chi(u-s)],
	\end{align}
where $\Theta$ is the Heaviside function. 
The corresponding expression for a scalar field with a derivative coupling
was obtained as (2.9) in~\cite{Juarez-Aubry:2014jba}. 

Asymptotic properties of $\F[NW]$ may be extracted from  
\eqref{FNW} by standard techniques. 
In particular, in the large gap limit, 
$|\omega| \to \infty$, 
the second term in \eqref{FNW} has falloff $\BigOinf{|\omega|^{-1}}$, 
as is seen by integration by parts, integrating the trigonometric factor~\cite{Wong:2001}.

\subsection{Eternal MIT Wall} \label{Section Static MIT Wall}

Consider next an eternal MIT wall. 
Using the results of a previous analysis in~\cite{WanMokhtar:2018lwi}, we show in Appendix \ref{Appendix Static MIT Wall Response Function} that
\begin{align} \label{FMIT}
		\F[MIT] = \F[NW] + \F[MIT][(1)] + \F[MIT][(2)],
	\end{align}
where $\F[NW]$ is as given in \eqref{FNW} and the other terms are 
\begin{align}
		\F[MIT][(1)]
			&= \frac{1}{4\pi^2d} \text{Re} \bigg\{ e^{-2i\omega{d}}\bigg[ i\pi G(2d)
				\nonumber \\
			&\qquad \qquad \qquad
				+ \int_{0}^{\infty} \dd{r} \frac{G(2d+r)e^{-i\omega{r}}
					- G(2d-r)e^{i\omega{r}}}{r} \bigg] \bigg\}, \label{Fsub{MIT1}} \\
		\F[MIT][(2)]
			&= \frac{1}{2\pi^2d^2} \int_{0}^{\infty} \dd{s} \cos(\omega{s}) \, G(s) , 
\label{Fsub{MIT2}}
\end{align}
where
\begin{align} 
G(y) = \int_{-\infty}^{\infty} \dd{u} \chi(u)\chi(u-y). 
\label{G(y)}
\end{align}
We also show in Appendix \ref{Appendix Static MIT Wall Response Function} 
that the large gap asymptotic form of $\F[MIT]$ is 
\begin{align}
\label{F{MIT} large energy gap limit}
		\F[MIT]
			= -\frac{\omega\Theta(-\omega)}{\pi} \int_{-\infty}^{\infty} \dd{u} [\chi(u)]^2
				+ \frac{\Theta(-\omega)\sin(2\omega{d})}{2\pi{d}} G(2d)
				+ \BigOinf{|\omega|^{-1}} . 
\end{align}

For the normally-ordered interaction $H_{\text{int}}^{\text{no}}$~\eqref{Normally ordered scalar density interaction Hamiltonian}, 
$\F[MIT][(2)]$ is absent from~\eqref{FMIT}, 
but the large gap asymptotic form 
is still given by~\eqref{F{MIT} large energy gap limit}.

\subsection{General Evolving Wall} \label{Section General Evolving Wall}

Consider now a general evolving wall.
Let 
\begin{align}
		\chi_{1}(\tau)
			&= \cos\bigl(\varphi(\tau)+\varphi_{1}(\tau)\bigr)\cos\theta(\tau)\chi(\tau),
				\label{GEW Modified Switch 1} \\
		\chi_{2}(\tau)
			&= \sin\bigl(\varphi(\tau)+\varphi_{1}(\tau)\bigr)\cos\theta(\tau)\chi(\tau),
				\label{GEW Modified Switch 2} \\
		\chi_{3}(\tau)
			&= \cos\bigl(\varphi(\tau)+\varphi_{2}(\tau)\bigr)\sin\theta(\tau)\chi(\tau),
				\label{GEW Modified Switch 3} \\
		\chi_{4}(\tau)
			&= \sin\bigl(\varphi(\tau)+\varphi_{2}(\tau)\bigr)\sin\theta(\tau)\chi(\tau), 
				\label{GEW Modified Switch 4}
\end{align}
and let 
\begin{align}
G_{j}(y) = \int_{-\infty}^{\infty} \dd{u} \chi_{j}(u)\chi_{j}(u-y) . 
\label{eq:Gjy}
\end{align}
In words, $\chi_j$ modulate the switching function $\chi$ by the wall evolution, 
and $G_j$ are the corresponding generalisations of~\eqref{G(y)}. 
With this notation, 
we show in Appendix \ref{Appendix Evolving Wall} that 
\begin{align} \label{Detector Response to GEW}
		\F[GEW] = \F[GEW][(0)] + \F[GEW][(1)] + \F[GEW][(2)],
\end{align}
where the subscript ``GEW'' indicates a general evolving wall,
	\begin{align}
		\F[GEW][(0)]
			&= -\frac{\omega\Theta(-\omega)}{\pi} \int_{-\infty}^{\infty} \dd{u} [\chi(u)]^2
					+ \Delta\F[GEW][(0)],
						\label{Fsub{GEW}0 Modified Regulator-free} \\
		\F[GEW][(1)]
			&= \frac{\Theta(-\omega)\sin(2\omega{d})}{2\pi{d}} 
					[G_{3}(2d) - G_{4}(2d)] + \Delta\F[GEW][(1)],
						\label{Fsub{GEW}1 Modified Regulator-free} \\
		\F[GEW][(2)]
			&= \frac{1}{2\pi^2d^2} \int_{0}^{\infty} \dd{s} \cos(\omega{s}) G_{3}(s),
						\label{Fsub{GEW}2 Modified Regulator-free}
\end{align}
and 
\begin{align}
		\Delta\F[GEW][(0)]
			&=  \frac{1}{\pi^2} \sum_{j = 1}^{4}
					\int_{0}^{\infty} \dd{s} \frac{\cos\omega{s}}{s^2}
					\int_{-\infty}^{\infty} \dd{u} \chi_{j}(u)[\chi_{j}(u)-\chi_{j}(u-s)],
						\label{DeltaFsub{GEW}0} \\
		\Delta\F[GEW][(1)]
			&= \frac{1}{4\pi^2d} \bigg[ \cos(2\omega{d})
					\int_{0}^{\infty} \dd{r} \cos\omega{r} \, 
						\frac{G_{3}(2d+r) - G_{3}(2d-r)}{r}
							\nonumber \\
			&\quad
				- \cos(2\omega{d})
					\int_{0}^{\infty} \dd{r} \cos\omega{r} \, 
						\frac{G_{4}(2d+r) - G_{4}(2d-r)}{r}
							\nonumber \\
			&\quad
				- \sin(2\omega{d})
					\int_{0}^{\infty} \dd{r} \sin\omega{r} \, 
						\frac{G_{3}(2d+r) + G_{3}(2d-r) - 2G_{3}(2d)}{r}
							\nonumber \\
			&\quad
				+ \sin(2\omega{d})
					\int_{0}^{\infty} \dd{r} \sin\omega{r} \, 
						\frac{G_{4}(2d+r) + G_{4}(2d-r) - 2G_{4}(2d)}{r} \bigg] . 
						\label{DeltaFsub{GEW}1}
\end{align}
The large gap asymptotic form of $\F[GEW]$ is 
\begin{align} 
\label{Detector Response to GEW Large Energy Gap}
\F[GEW]
&= -\frac{\omega\Theta(-\omega)}{\pi} \int_{-\infty}^{\infty} \dd{u} [\chi(u)]^2
%\nonumber \\
%&\quad 
+ \frac{\Theta(-\omega) \sin(2\omega{d})}{2\pi{d}}
[G_{3}(2d) - G_{4}(2d)] 
%\nonumber \\
%&\quad
+ \BigOinf{|\omega|^{-1}}, 
\end{align}
which may be found as in the passage from 
\eqref{FMIT}
to 
\eqref{F{MIT} large energy gap limit} for $\F[MIT]$; the key observation 
is that the functions $\chi_{j}$ \eqref{GEW Modified Switch 1}--\eqref{GEW Modified Switch 4} 
are smooth, including the parametrisation coordinate singularities at 
$\theta=0$ and $\theta=\pi/2$, and they have compact support. 

For $H_{\text{int}}^{\text{no}}$~\eqref{Normally ordered scalar density interaction Hamiltonian}, 
$\F[GEW][(2)]$ is absent from~\eqref{Detector Response to GEW}, 
but the large gap asymptotic form 
is still given by~\eqref{Detector Response to GEW Large Energy Gap}.

	\subsection{Wall Creation} \label{Section Wall Appearance}
	
We are now ready to specialise to the creation of the MIT wall through the scenario described in Section \ref{Section Wall Appearance and Disappearance}. 

Using \eqref{Discrete Family of Boundary Conditions}, 
the modulated switching functions \eqref{GEW Modified Switch 1}--\eqref{GEW Modified Switch 4} read
\begin{align}
		\chi_{1}(\tau)
			&= \cos\bigl(2n\theta(\tau)\bigr)\cos\theta(\tau)\chi(\tau), \\
		\chi_{2}(\tau)
			&= \sin\bigl(2n\theta(\tau)\bigr)\cos\theta(\tau)\chi(\tau), \\
		\chi_{3}(\tau)
			&= (-1)^{n+1}\cos\bigl(2n\theta(\tau)\bigr)\sin\theta(\tau)\chi(\tau), 
\label{eq:creation-chi3}\\		
		\chi_{4}(\tau)
			&= \sin\bigl(2n\theta(\tau)\bigr)\sin\theta(\tau)\chi(\tau).
\label{eq:creation-chi4}
\end{align}
For the wall creation profile \eqref{Condition of Theta for Wall Appearance}, this gives 
	\begin{align}
		\chi_{1}(\tau) &=
			\begin{cases}
				\chi(\tau) & \mkern+68mu \text{for } \tau \leq t_{0}, \\
				\cos\bigl(2nh_{1}(\tau)\bigr)\cos h_{1}(\tau)\chi(\tau) & \mkern+68mu \text{for } t_{0} < \tau < t_{0} + \delta, \\
				0 & \mkern+68mu \text{for } \tau \geq t_{0} + \delta,
			\end{cases} \label{Modified Switch 1 for Wall Appearance} \\
		\chi_{2}(\tau) &=
			\begin{cases}
				0 & \mkern+68mu \text{for } \tau \leq t_{0}, \\
				\sin\bigl(2nh_{1}(\tau)\bigr)\cos h_{1}(\tau)\chi(\tau) & \mkern+68mu \text{for } t_{0} < \tau < t_{0} + \delta, \\
				0 & \mkern+68mu \text{for } \tau \geq t_{0} + \delta,
			\end{cases} \label{Modified Switch 2 for Wall Appearance} \\
		\chi_{3}(\tau) &=
			\begin{cases}
				0 & \text{for } \tau \leq t_{0}, \\
				(-1)^{n+1}\cos\bigl(2nh_{1}(\tau)\bigr)\sin h_{1}(\tau)\chi(\tau) & \text{for } t_{0} < \tau < t_{0} + \delta, \\
				-\chi(\tau) & \text{for } \tau \geq t_{0} + \delta,
			\end{cases} \label{Modified Switch 3 for Wall Appearance} \\
		\chi_{4}(\tau) &=
			\begin{cases}
				0 & \mkern+68mu \text{for } \tau \leq t_{0}, \\
				\sin\bigl(2nh_{1}(\tau)\bigr)\sin h_{1}(\tau)\chi(\tau) & \mkern+68mu \text{for } t_{0} < \tau < t_{0} + \delta, \\
				0 & \mkern+68mu \text{for } \tau \geq t_{0} + \delta.
			\end{cases} \label{Modified Switch 4 for Wall Appearance}
	\end{align}
For $n=0$, $\chi_2$ and $\chi_4$ vanish everywhere, whereas 
$\chi_1$ interpolates between $\chi$ and $0$, and $\chi_3$ interpolates between $0$ and $-\chi$, over the interval $t_{0} < \tau < t_{0} + \delta$, in which the wall is created. 
For $n\ne0$, the behaviour changes only in the interval $t_{0} < \tau < t_{0} + \delta$, where all four $\chi_j$ now oscillate, the number of oscillations increasing with~$|n|$. 

The detector's response is obtained by inserting 
\eqref{Modified Switch 1 for Wall Appearance}--\eqref{Modified Switch 4 for Wall Appearance}
into \eqref{eq:Gjy}--\eqref{Detector Response to GEW Large Energy Gap}. 
We denote this response by 
$\F[WC]$, where ``WC'' stands for wall creation. 

Our main interest is in the limit of rapid wall creation, $\delta\to0$. 
We start with two preparatory observations. 

First, if the detector operates only within the future lightcone of the fully developed wall, 
we have $\chi(\tau) = 0$ for $\tau \leq t_{0} + \delta$, 
whence $\chi_{1}(\tau) = \chi_{2}(\tau) = \chi_{4}(\tau) = 0$ and $\chi_{3}(\tau) = -\chi(\tau)$. 
It follows that the response is identical to that for an eternal MIT wall: 
$\F[WC] = \F[MIT]$. 
The creation of the wall, no matter how rapid, leaves no lingering trace in the response of a detector that operates only within the lightcone of the fully developed wall. 
This was to be expected, as the field is massless, propagating strictly along null rays. 

Second, consider the large $|\omega|$ asymptotic form of~$\F[WC]$, 
obtained from \eqref{Detector Response to GEW Large Energy Gap} 
with $\chi_3$ and $\chi_4$ in \eqref{eq:creation-chi3} and~\eqref{eq:creation-chi4}. 
When $\delta < 2d$, we have 
$G_4(2d) =0$ and 
\begin{align}
G_3(2d) &= \int_{t_{0}}^{\infty} \dd{u} \chi(u)\chi(u+2d) 
\notag \\ 
& \hspace{3ex}
- \int_{t_0}^{t_0+\delta} \dd{u} \chi(u)\chi(u+2d) 
\left[ 1 + (-1)^{n+1}\cos\bigl(2nh_{1}(u)\bigr)\sin h_{1}(u)\right] , 
\end{align}
where the first term is independent of $\delta$ 
and the second term is of order $\BigO{\delta}$ as $\delta\to0$. 
This shows that the leading and next-to-leading terms 
in the large $|\omega|$ asymptotic form of $\F[WC]$ 
remain finite in the $\delta\to0$ limit: 
if $\F[WC]$ has a divergence as $\delta\to0$, 
the divergence does not show up in the two leading terms in the 
large $|\omega|$ expansion of~$\F[WC]$. 

Now, consider the issue of main interest, the limit $\delta\to0$ with everything else fixed. 
We show in Appendix \ref{Appendix Wall Appearance} that in this limit we have 
\begin{align} 
\F[WC] &= -\frac{[\chi(t_{0})]^2}{\pi^2} \ln(\mu\delta) 
+ \Fconst[WC][(c)] + \F[NW] + \F[RWC][(0)]
+ \F[RWC][(1)] + \F[RWC][(2)] 
\notag\\
&\qquad + \BigO{\delta\ln\delta},
\label{F{RWC}}
\end{align}
where $\F[NW]$ is given by~\eqref{FNW}, 
	\begin{align}
		\Fconst[WC][(c)]
			&= - \frac{\chi(t_{0})}{\pi^2}\int_{t_{0}}^{\infty} \dd{u} \chi'(u)\ln\bigl(\mu(u-t_{0})\bigr) 
				- \frac{1}{\pi^2} \int_{t_{0}}^{\infty} \dd{u} \chi(u)
					\int_{u-t_{0}}^{\infty} \dd{s} \frac{\chi'(u-s)}{s} , 
\label{Fsub{WC}c} \\
\F[RWC][(0)]
			&= -\frac{1}{\pi^2} \int_{0}^{\infty} \dd{s} \frac{1-\cos(\omega{s})}{s^2} f(s), 
						\label{Fsub{RWC}0} \\
		\F[RWC][(1)]
			&= \frac{\Theta(-\omega)}{2\pi{d}}
					\sin(2\omega{d}) H_{1}(2d) \nonumber \\
			&\quad
				+ \frac{1}{4\pi^2d} \cos(2\omega{d})
					\int_{0}^{\infty} \dd{r} \cos\omega{r} \, 
						\frac{H_{1}(2d+r) - H_{1}(2d-r)}{r} \nonumber \\
			&\quad
				- \frac{1}{4\pi^2d} \sin(2\omega{d})
					\int_{0}^{\infty} \dd{r} \sin\omega{r} \, 
						\frac{H_{1}(2d+r) + H_{1}(2d-r) - 2H_{1}(2d)}{r} \nonumber \\
			&\quad
				+ \frac{1}{4\pi^2d} \cos(2\omega{d})
					\int_{2d}^{\infty} \dd{r} \frac{\cos\omega{r}}{r} f(r-2d)
						\nonumber \\
			&\quad
				+ \frac{1}{4\pi^2d} \sin(2\omega{d})
					\int_{2d}^{\infty} \dd{r} \frac{\sin\omega{r}}{r} f(r-2d),
						\label{Fsub{RWC}1} \\
		\F[RWC][(2)]
			&= \frac{1}{2\pi^2d^2} \int_{0}^{\infty} \dd{s} \cos(\omega{s}) H_{1}(s),
					\label{Fsub{RWC}2}
\end{align}
with 
\begin{align}
		f(s) &= \int_{t_{0}- (s/2)}^{t_{0}+ (s/2)} \dd{v}
				\chi\bigl(v+ \tfrac12 s\bigr)\chi\bigl(v- \tfrac12 s\bigr), \label{f(s)} \\
		H_{1}(y) &= \int_{t_{0}}^{\infty} \dd{u} \chi(u)\chi(u+y), 
				\label{H_{1}(y)}
\end{align}
and $\mu$ is a positive constant of dimension inverse length. 
$\mu$~is introduced for dimensional consistency in the first two terms in~\eqref{F{RWC}}, 
but it cancels out from the sum of these two terms. 
Note that these two terms contribute to $\F[WC]$ only by an $\omega$-independent additive constant. 

For the normally-ordered interaction Hamiltonian~\eqref{Normally ordered scalar density interaction Hamiltonian}, 
the only difference is that $\F[RWC][(2)]$ \eqref{Fsub{RWC}2} is absent. 

The key outcome from \eqref{F{RWC}} is that the detector's response $\F[WC]$ diverges in the $\delta\to0$ limit 
if and only if $\chi(t_0)\ne0$, that is, if and only if the detector operates when it crosses the light cone of the wall creation event. 
While this is qualitatively similar to the divergence in the energy expectation value found in Section~\ref{Section Wall Appearance and Disappearance Stress Energy Tensor}, there are two quantitative differences. First, the divergence in the detector's response is weaker, 
only logarithmic in $\delta$, while the divergence in the energy was proportional to $1/\delta$. 
Second, the divergence in the detector's response is independent of the details of the wall creation profile function $h_1$ introduced in~\eqref{Condition of Theta for Wall Appearance}, and of the integer parameter $n$ in the boundary condition family~\eqref{Discrete Family of Boundary Conditions}; the divergence in the energy, by contrast, depends on both of these, as found in Section~\ref{Section Wall Appearance and Disappearance Stress Energy Tensor}. 

We also emphasise that the divergence in $\F[WC]$ \eqref{F{RWC}} is independent of the gap~$\omega$. 
Subtracting the $\omega$-independent divergence from \eqref{F{RWC}} and taking the limit $\delta\to0$ leaves 
a finite remainder with nontrivial $\omega$-dependence, given by $\F[NW]$ and $\F[RWC][(0)]$, $\F[RWC][(1)]$ and $\F[RWC][(2)]$ in~\eqref{F{RWC}}. The large $|\omega|$ asymptotics of this remainder can be investigated by standard techniques; 
however, this would only address the situation in which $\delta\to0$ is taken first and $|\omega|\to\infty$ subsequently, 
and would not address limits in which the smallness of $\delta$ is related to largeness of~$|\omega|$.

Finally, the divergence in our detector's response is in a sharp contrast with the response of a detector coupled linearly to the time derivative of a scalar field, which remains finite in the rapid wall creation limit~\cite{Brown:2015yma}. The scalar field case however requires an infrared regulator, and the response diverges if the infrared regulator is removed. 
Our fermion field detector requires no infrared regulator, 
and may hence be a better model for processes in which a wall is created by fields with dynamics of their own.

\subsection{Wall Demolition} \label{Section Wall Disappearance}

For the demolition of the MIT wall through the scenario described in Section~\ref{Section Wall Appearance and Disappearance}, 
the techniques of Section \ref{Section Wall Appearance} and Appendix \ref{Appendix Wall Appearance} are adaptable in a straightforward way, 
and the results are very similar. We shall here only outline the key points. We denote the response function by $\F[WD]$. 

First, if the detector operates only within the future lightcone of the fully demolished wall, 
the response is identical to that in Minkowski vacuum: $\F[WD] = \F[NW]$. 

Second, the leading and next-to-leading terms 
in the large $|\omega|$ asymptotic form of $\F[WD]$ 
remain finite in the $\delta\to0$ limit. 

Third, and for the issue of main interest, in the limit of $\delta\to0$ with everything else fixed, 
$\F[WD]$ has a formula similar to $\F[WC]$ \eqref{F{RWC}}, 
but in the $\omega$-dependent terms the function 
$H_1(y)$ \eqref{H_{1}(y)} is replaced by 
\begin{align}
H_{2}(y)
:= \int_{-\infty}^{t_{0}} \dd{u} \chi(u)\chi(u-y).
\end{align}
The divergence considerations as $\delta\to0$ are hence the same as for wall creation. 
		
	\section{Discussion} \label{Section Discussion}

We have analysed the energetic cost of building and demolishing a wall for a 
massless Dirac field in $(1+1)$-dimensional Minkowski spacetime, 
and the response of an Unruh-DeWitt particle detector to the generated radiation. 
The wall was mechanically static, at $z=0$ in a set of standard Minkowski coordinates $(t,z)$; 
however, its reflection and transmission properties followed a prescribed time evolution, 
implemented by a smoothly time-dependent self-adjoint extension of the Dirac Hamiltonian on the half-lines 
of positive and negative $z$ joined at the distinguished point $z=0$. 
The field was initially prepared in an ``in-vacuum'', 
corresponding to no radiation towards the wall from the infinity. 

We first addressed a general wall evolution. 
The renormalised energy density, obtained by point-splitting regularisation, 
was shown to be well defined, and given by 
\eqref{General Time-Dependent Vacuum Energy Density Expectation Value Right}
and~\eqref{General Time-Dependent Vacuum Energy Density Expectation Value Left}. 
The response of an inertial Unruh-DeWitt detector, 
moving parallelly to the wall, and treated to first order in perturbation theory, 
was also shown to be well defined, and given by~\eqref{Detector Response to GEW}; 
the wall modulates the 
detector's switching function by time-dependent factors determined by the evolution of the wall. 

We then specialised to wall evolutions that describe the 
creation or demolition, over a finite time interval, 
of an impermeable wall that enforces the MIT bag boundary condition on both sides of the wall. 
In the $\mathrm{U}(2)$ group of boundary conditions that characterise the self-adjoint extensions, we identified 
a countable set of one-parameter subgroups that mediate the creation or demolition, 
and we analysed the asymptotic behaviour of the energy and the detector 
response in the limit of rapid creation or demolition. 

The main outcome was that, in the limit of rapid creation or demolition of the wall, 
the total energy emitted into the field diverges, and so does the detector's response, but the rates of divergence differ: 
denoting the duration of the creation or demolition by~$\delta$, the emitted energy diverges (at least) 
proportionally to~$\delta^{-1}$, but the detector's response diverges only logarithmically in~$\delta$. 
This result is consistent with earlier evidence \cite{Brown:2015yma,Louko:2014aba,Martin-Martinez:2015dja} to the effect 
that a localised matter system coupled 
to the field is less sensitive to a rapid wall creation or demolition than energetic considerations would suggest. 
We also obtained some technical results on the choice of the one-parameter subgroup in $\mathrm{U}(2)$ along which the creation or demolition takes place, and on the asymptotic large gap behaviour of the detector's response. 

Our analysis contained a number of limitations that suggest questions for future work. 
First, to step from $1+1$ dimensions to our home spacetime dimension of $3+1$, 
our boundary condition techniques are adaptable to a mechanically static spherical wall in 
$3+1$ dimensions, by separation of variables, in line with the scalar field wall considered in~\cite{Carrington:2018ikq}, 
but with the additional complication that the radial equation contains also a potential term. 
Second, we chose the coupling between the detector and the Dirac field for mathematical simplicity; 
it should be clarified how well this coupling models localised matter interactions that are known exist in nature, 
as has been discussed for a detector coupled to a scalar field in~\cite{MartinMartinez:2012th,Alhambra:2013uja}. 
Third, the boundary conditions at the wall were characterised by self-adjoint extensions of the Dirac Hamiltonian, with a well-known mathematical theory, but --- perhaps with the exception of the MIT wall boundary conditions --- without a clear physical motivation, and certainly without dynamics of its own. Could the wall be given internal dynamics, perhaps building on the idealised delta-potential of Section \ref{Section Wall Appearance and Disappearance} and Appendix \ref{Appendix Delta Potential}, or by introducing fields that live on the boundary, perhaps by the techniques of~\cite{G.:2015yxa,Barbero:2017kvi,Dappiaggi:2018pju}? 
We leave these topics subject to future work.

\acknowledgments{We thank Alex Schenkel for helpful discussions and Lee Sangchul for pointing out the use of Cauchy-Schwarz inequality to derive \eqref{Wall creation or demolition Cauchy-Schwarz inequality}.
WMHWM was supported by the Ministry of Higher Education (MOHE) 
Malaysia and Universiti Sains Malaysia (USM). 
JL was supported in part by 
United Kingdom Research and Innovation (UKRI) Science and 
Technology Facilities Council (STFC) grant ST/S002227/1 ``Quantum Sensors for Fundamental Physics'' 
and by Theory Consolidated Grants ST/J000388/1 and ST/P000703/1.}

\appendix
	
\section{Appendix: Boundary Condition} \label{Appendix Boundary Condition}

In this appendix we verify the Dirac operator self-adjoint extensions results stated in Section \ref{Section General Boundary Condition}.

\subsection{Self-adjoint Extensions of $H = -\imaginary\alpha\partial_{z}$ on a Disjoint Pair of Half-Lines} \label{Appendix Self-adjoint Extensions}

In the notation of Section~\ref{Section General Boundary Condition}, 
we consider the massless Dirac operator $H = -i\alpha\partial_{z}$ on the Hilbert space $\mathcal{H} = \overset{{}_{\leftarrow}}{\mathcal{H}} \oplus \overset{{}_{\rightarrow}}{\mathcal{H}}$, where $\overset{{}_{\leftarrow}}{\mathcal{H}} = L_2((-\infty,0]) \oplus L_2((-\infty,0])$ and $\overset{{}_{\rightarrow}}{\mathcal{H}} = L_2([0,\infty)) \oplus L_2([0,\infty))$.
We write the inner product in $\mathcal{H}$ as 
\begin{align} \label{Inner Product Punctured}
		(\phi_{1},\phi_{2})
			= \int_{-\infty}^{0} [\phi_{1}(z)]^{\dagger} \phi_{2}(z) \dd{z}
				+ \int_{0}^{\infty} [\phi_{1}(z)]^{\dagger} \phi_{2}(z) \dd{z},
\end{align}
where $\phi_{1},\phi_{2} \in \mathcal{H}$. 

To begin with, let the domain of $H$ be 
\begin{align} \label{Initial domain of H}
		\mathcal{D}(H)
			= \{\phi \in \mathcal{H}
				\mid \partial_{z}\phi \in \mathcal{H},
				\phi(0_{-}) = \phi(0_{+}) = 0 \},
\end{align}
where $\phi(0_{-}) := \lim_{z\to0_-} \phi(z)$ and $\phi(0_{+}) := \lim_{z\to0_+} \phi(z)$. 
These limiting values exist because the condition 
$\partial_{z}\phi \in \mathcal{H}$ makes the projections of $\phi$ to $\overset{{}_{\leftarrow}}{\mathcal{H}}$ and $\overset{{}_{\rightarrow}}{\mathcal{H}}$ absolutely continuous, and hence in particular continuous, on respectively $(-\infty,0]$ and $[0,\infty)$. 

Note also that for every $\phi \in \mathcal{H}$ such that $\partial_{z}\phi \in \mathcal{H}$, 
it holds (\cite{Richtmyer:1978Vol1}, pages 85--86) that 
\begin{align} \label{Value at Infinity}
\lim_{z \to -\infty} \phi(z)
= 0 = \lim_{z \to \infty} \phi(z) . 
\end{align}
In particular, \eqref{Value at Infinity} holds for every $\phi \in \mathcal{D}(H)$. 

Given these preparations, consider the adjoint of~$H$, denoted~$H^{\dagger}$. 
The domain of $H^{\dagger}$, denoted by $\mathcal{D}(H^{\dagger})$, 
consists of all $\psi\in \mathcal{H}$ such that the map $\phi \mapsto (\psi,H\phi)$ 
is continuous, 
and $H^{\dagger}$ is then defined by the property that 
\begin{align} \label{Definition of Adjoint 1}
(\psi,H\phi) = (H^{\dagger}\psi,\phi) 
% \qquad \forall \phi \in \mathcal{D}(H).
\end{align}
for all $\phi \in \mathcal{D}(H)$~\cite{Bonneau:1999zq}. 
Using $H = -i\alpha\partial_{z}$, 
integration by parts in \eqref{Definition of Adjoint 1} gives 
\begin{align} \label{Definition of Adjoint 2}
(-i\alpha\partial_{z}\psi,\phi) = (H^{\dagger}\psi,\phi) 
% \qquad \forall \phi \in \mathcal{D}(H),
\end{align}
for all $\phi \in \mathcal{D}(H)$, possibly subject to boundary conditions on~$\psi$; 
however, by continuity of the map $\phi \mapsto (\psi,H\phi)$, 
\eqref{Definition of Adjoint 2} 
shows that $\partial_z \psi \in \mathcal{H}$, 
whence $\lim_{z\to 0_{\pm}} \psi(z)$ exist and $\lim_{z\to \pm\infty} \psi(z) =0$, 
so that \eqref{Definition of Adjoint 2} holds without boundary conditions on~$\psi$.  
Hence $H^\dagger = -i\alpha\partial_{z}$ and 
\begin{align}
\mathcal{D}(H^{\dagger})
= \{\psi \in \mathcal{H}
\mid \partial_{z}\psi \in \mathcal{H} \}.
\end{align}

Since $\mathcal{D}(H) \subsetneq \mathcal{D}(H^{\dagger})$, $H$ with the domain $\mathcal{D}(H)$ is not self-adjoint. To find the self-adjoint extensions of $(H,\mathcal{D}(H))$, we need to enlarge 
$\mathcal{D}(H)$ and shrink $\mathcal{D}(H^{\dagger})$ 
until the two coincide. 

To this end, let $\mathcal{D}'(H)$ be an enlargement of $\mathcal{D}(H)$, 
where we still must have $\partial_z \phi \in \mathcal{H}$ for every $\phi \in \mathcal{D}'(H)$, 
from which it follows that $\lim_{z\to 0_{\pm}} \phi(z)$ exist and $\lim_{z\to \pm\infty} \phi(z) =0$; 
the only remaining freedom in $\mathcal{D}'(H)$ is hence a possible linear boundary condition for 
$\lim_{z\to 0_{\pm}} \phi(z)$. 
Let $\mathcal{D}'(H^{\dagger})$ be the corresponding domain of~$H^{\dagger}$: by the above observations 
it follows that $\partial_z \psi \in \mathcal{H}$ for every $\psi \in \mathcal{D}'(H^\dagger)$, 
from which it follows that  
$\lim_{z\to 0_{\pm}} \psi(z)$ exist and $\lim_{z\to \pm\infty} \psi(z) =0$, and there may be a linear boundary condition for $\lim_{z\to 0_{\pm}} \psi(z)$, determined by that in $\mathcal{D}'(H)$. 

For $\phi \in \mathcal{D}'(H)$ and $\psi \in \mathcal{D}'(H^\dagger)$, 
we then find 
	\begin{align} \label{Sesquilinear}
		B(\psi,\phi)
			&:=	i\left[ (\psi, H\phi) - (H^\dagger\psi,\phi) \right] \nonumber \\
			&= \psi^{\dagger}(0_{-})\alpha\phi(0_{-}) - \psi^{\dagger}(0_{+})\alpha\phi(0_{+}),
	\end{align}
by integration by parts. 
$H$ is hence self-adjoint iff the linear boundary conditions in 
$\mathcal{D}'(H)$ and $\mathcal{D}'(H^{\dagger})$ coincide and $B(\psi,\phi)$ is vanishing. 

Expanding $\phi$ and $\psi$ in the spinor basis $\{U_{+},U_{-}\}$, 
\eqref{Sesquilinear} gives 
	\begin{align}
		B(\psi,\phi) =
			u^\dagger
			\begin{pmatrix}
				I & 0 \\
				0 & -I
			\end{pmatrix}
			v,
	\end{align}
	where $I$ is the $2 \times 2$ identity matrix and
	\begin{align}
		u := 
			\begin{pmatrix}
				\psi_{+}(0_{-}) \\
				\psi_{-}(0_{+}) \\
				\psi_{+}(0_{+}) \\
				\psi_{-}(0_{-})
			\end{pmatrix}, \qquad
		v := 
			\begin{pmatrix}
				\phi_{+}(0_{-}) \\
				\phi_{-}(0_{+}) \\
				\phi_{+}(0_{+}) \\
				\phi_{-}(0_{-})
			\end{pmatrix}.
	\end{align}
It then follows (for the spelled-out argument, see Appendix A of~\cite{Louko:2014oha}) 
that the self-adjoint extensions of $(H,\mathcal{D}(H))$ 
are those in which $\mathcal{D}'(H)$ has the boundary condition 
\begin{align}
		\begin{pmatrix}
			\phi_{+}(0_{+}) \\
			\phi_{-}(0_{-})
		\end{pmatrix}
		&=
		U
		\begin{pmatrix}
			\phi_{+}(0_{-}) \\
			\phi_{-}(0_{+})
		\end{pmatrix}, \label{Dirac operator self-adjoint extensions boundary condition 1}
	\end{align}
where the matrix $U \in \mathrm{U}(2)$ specifies the extension.

\subsection{Boundary condition in terms of a $\delta$-potential} \label{Appendix Delta Potential}

We now show that some of the boundary conditions \eqref{Dirac operator self-adjoint extensions boundary condition 1} may be described in terms of a Dirac operator that involves a $\delta$-function potential at $z=0$, of the form 
\begin{align} \label{Dirac Equation Delta Potential General appendix}
H_\delta
:= 
-\imaginary\alpha\partial_{z} + \bigl(S\beta+V\bigr)\delta(z) , 
\end{align}
where the real-valued constants $S$ and $V$ are determined by the matrix $U$ in \eqref{Dirac operator self-adjoint extensions boundary condition 1} and the action of the delta-function on a discontinuous function is defined to be 
\begin{align} 
\lim_{\epsilon \to 0_{+}} \int_{-\epsilon}^{+\epsilon} \delta(z)\phi(z) \dd{z}
:= \frac{1}{2}\bigl(\phi(0_{-}) + \phi(0_{+})\bigr) . 
\label{Delta Potential Common Definition appendix}
\end{align}	
Dirac operators of this form can be recovered as a limit of a non-distributional potential if the potential is made suitably non-local~\cite{Calkin:1988zz,Mckellar:1987ru,Calkin:1987}. 

Given 
\eqref{Dirac Equation Delta Potential General appendix}--\eqref{Delta Potential Common Definition appendix}, the condition that $H_\delta \phi$ have no distributional part at $z=0$ reads 
\begin{align}
-\imaginary\alpha \bigl(\phi(0_{+}) - \phi(0_{-})\bigr)
+ 
\frac12 \bigl(S\beta+V\bigr) \bigl(\phi(0_{+}) + \phi(0_{-})\bigr) = 0 . 
\label{eq:deltapotential-condition}
\end{align}
Requiring \eqref{eq:deltapotential-condition} to be equivalent to \eqref{Dirac operator self-adjoint extensions boundary condition 1}, we obtain four linear equations for the four components of the matrix~$U$, with the unique solution 
\begin{align}
U_{11} = U_{22} & = \frac{4 + V^2 - S^2}{(2 + i V)^2 + S^2} , 
\label{eq:Udiag-deltapot-sol}
\\
U_{12} = U_{21} & = - \frac{4iS}{(2 + i V)^2 + S^2} . 
\label{eq:Uoffdiag-deltapot-sol}
\end{align}
The conditions $U_{11} = U_{22}$ and $U_{12} = U_{21}$ imply that 
\begin{align} 
		U =
		\exp{\imaginary\tilde\varphi}
		\begin{pmatrix}
			\cos\tilde\theta
				& i \sin\tilde\theta \\
			i \sin\tilde\theta
				& \cos\tilde\theta
		\end{pmatrix}, 
\label{eq:Upar-tilded}
\end{align}
where $\tilde\varphi$ and $\tilde\theta$ are each periodic with period $2\pi$ and the only redundancy in the parametrisation is that $(\tilde\varphi, \tilde\theta) \sim (\tilde\varphi + \tilde\pi, \tilde\theta + \pi)$. 
If $\cos\tilde\theta + \cos\tilde\varphi \ne0$, the pair 
\eqref{eq:Udiag-deltapot-sol}--\eqref{eq:Uoffdiag-deltapot-sol} can then be solved for 
$S$ and~$V$, with the outcome 
\begin{align}
S = 
- \frac{2 \sin\tilde\theta}{\cos\tilde\theta + \cos\tilde\varphi} , 
\hspace{5ex}
V = 
- \frac{2 \sin\tilde\varphi}{\cos\tilde\theta + \cos\tilde\varphi} . 
\end{align}

The one-parameter subgroups \eqref{Discrete Family of Boundary Conditions} introduced in Section \ref{Section Wall Appearance and Disappearance} are in the family \eqref{eq:Upar-tilded} with 
$\tilde\theta = (-1)^{n+1} \theta$ and $\tilde\varphi = 2 n \theta$, yielding for $S$ and $V$ the formulas~\eqref{S(t) and V(t)}.

\section{Appendix: Particle Detector} \label{Appendix Particle Detector}

In this appendix we verify the particle detector response results stated in 
Sections \ref{Section Static MIT Wall}--\ref{Section Wall Disappearance}. 

As preparation, recall that the response function is given by~\eqref{Detector Response Function}, 
where the two-point function $W^{(2,\bar{2})}(\tau,\tau')$ is a distribution, 
represented by the $\epsilon\to0_+$ limit of a one-parameter family of functions~$W_\epsilon^{(2,\bar{2})}(\tau,\tau')$, 
as described in 
Section~\ref{subsec:responsefunction}. 
Recall also that
$W^{(2,\bar{2})}(\tau,\tau')^* = W^{(2,\bar{2})}(\tau',\tau)$ by construction, 
and this property is respected by the $\epsilon$-regularisation, 
$W_\epsilon^{(2,\bar{2})}(\tau,\tau')^* = W_\epsilon^{(2,\bar{2})}(\tau',\tau)$. 
It follows that we can write \eqref{Detector Response Function} as 
\begin{align} \label{Causal response function}
		\F = 2 \lim_{\epsilon\to0_+}\int_{-\infty}^{\infty} \dd{r} \chi(r) \int_{0}^{\infty} \dd{s} \chi(r-s)
				\Re{e^{- i \omega s} W_\epsilon^{(2,\bar{2})}(r,r-s)} , 
\end{align}
which will be convenient in what follows. 

% From now on we suppress the ``lim'' in the formulas. 

\subsection{Eternal MIT Wall} \label{Appendix Static MIT Wall Response Function}

For the eternal MIT wall at $z = 0$, the regularised two-point function 
$W_{\epsilon,\text{MIT}}^{(2,\bar{2})}$ is given for $z, z' > 0$ by \cite{WanMokhtar:2018lwi}
	\begin{align} \label{Static MIT wall two-point function}
		W_{\epsilon,\text{MIT}}^{(2,\bar{2})}(u,v;u',v')
			&= - \frac{1}{2\pi^{2}} \bigg[ \frac{ 1 }{ ( u - u' - i \epsilon ) ( v - v' - i \epsilon ) }
				- \frac{ 1 }{ ( u - v' - i \epsilon ) ( v - u' - i \epsilon ) }
					\nonumber \\
			&\qquad \qquad \qquad
				- 2 \frac{ u - v }{ ((u - v)^{2} + \epsilon^{2}) }
								\frac{ u' - v' }{ ((u' - v')^{2} + \epsilon^{2}) } \bigg],
	\end{align}
where the subscript ``MIT'' indicates the eternal MIT wall. 
The first two terms in \eqref{Static MIT wall two-point function} come from the first term in~\eqref{Vacuum two-point function in terms of propagators}, 
and the last term in 
\eqref{Static MIT wall two-point function} comes from the second term in~\eqref{Vacuum two-point function in terms of propagators}. Note that the first term in \eqref{Static MIT wall two-point function} is the two point function in Minkowski vacuum without a wall. 

For a detector on the trajectory~\eqref{Detector Trajectory}, 
substituting \eqref{Static MIT wall two-point function} into \eqref{Causal response function} gives
	\begin{align} \label{FMIT temporary}
		\F[MIT] = \F[NW] + \F[MIT][(1)] + \F[MIT][(2)],
	\end{align}
where $\F[NW]$, $\F[MIT][(1)]$ and $\F[MIT][(2)]$ come respectively from the first, second and third term in~\eqref{Static MIT wall two-point function}, 
and are given by 
\begin{align}
		\F[NW]
			&= \lim_{\epsilon \to 0_{+}} \frac{-1}{\pi^2}
				\int_{-\infty}^{\infty} \dd{u} \chi(u) \int_{0}^{\infty} \dd{s} \chi(u-s)
					\Re{\frac{e^{-i\omega{s}}}{(s-i\epsilon)^{2}}},
\label{eq:app:fnw}\\
		\F[MIT][(1)]
			&= \lim_{\epsilon \to 0_{+}} \frac{1}{\pi^2} \int_{-\infty}^{\infty} \dd{u} \chi(u) \int_{0}^{\infty} \dd{s} \chi(u-s)
					\Re{\frac{e^{-i\omega{s}}}{(s-i\epsilon)^{2}-4d^{2}}}, 
\label{eq:app:fmit1}\\
		\F[MIT][(2)]
			&= \frac{1}{2\pi^2d^2} \int_{-\infty}^{\infty} \dd{u} \chi(u)
				\int_{0}^{\infty} \dd{s} \chi(u-s) \cos(\omega{s}) , 
\label{eq:app:fmit2}
	\end{align}
where in \eqref{eq:app:fmit2} the $\epsilon\to0_+$ limit has already been taken. 

$\F[NW]$ \eqref{eq:app:fnw} is the Minkowski vacuum contribution. 
Taking the $\epsilon\to0_+$ limit gives \eqref{FNW} in the main text. 

In $\F[MIT][(2)]$ \eqref{eq:app:fmit2}, interchanging the integrals, 
justified by absolute convergence of the double integral, gives \eqref{Fsub{MIT2}} in the main text. 

In $\F[MIT][(1)]$~\eqref{eq:app:fmit1}, 
we use partial fractions to write 
\begin{align}
\F[MIT][(1)] = \F[MIT,s][(1)] + \F[MIT,r][(1)] , 
\end{align}
where 
\begin{align}
		\F[MIT,r][(1)]
			&= - \lim_{\epsilon \to 0_{+}} \frac{1}{4\pi^{2}d} \int_{-\infty}^{\infty} \dd{u} \chi(u) \int_{0}^{\infty} \dd{s} \chi(u-s)
					\Re{\frac{e^{-i\omega{s}}}{s + 2d - i\epsilon}}, \\
		\F[MIT,s][(1)]
			&= \lim_{\epsilon \to 0_{+}} \frac{1}{4\pi^{2}d} \int_{-\infty}^{\infty} \dd{u} \chi(u) \int_{0}^{\infty} \dd{s} \chi(u-s)
					\Re{\frac{e^{-i\omega{s}}}{s - 2d - i\epsilon}}. 
\end{align}
In $\F[MIT,r][(1)]$, the $\epsilon\to0_+$ limit can be taken under the integral, 
and a change of variables by $r = s + 2d$ and an interchange of the integrals gives 
\begin{align} \label{FMIT1r final}
\F[MIT,r][(1)]
&= -\frac{1}{4\pi^2d} \Re{ e^{-2i\omega{d}} \int_{2d}^{\infty}
\dd{r} \frac{G(2d-r) e^{i\omega{r}}}{r}}, 
\end{align}
where the function $G$ is given by~\eqref{G(y)}.
In $\F[MIT,s][(1)]$, whose integrand contains a singularity at $s = 2d$ when $\epsilon = 0$, 
we use the 
Sokhotsky formula, $\lim_{\epsilon \to 0_{+}} (x-\imaginary\epsilon)^{-1} = P(1/x) + \imaginary\pi\delta(x)$, where $P$ stands for the Cauchy principal value~\cite{Mukhanov:2007zz}. A further change of variables and interchange of the integrals gives 
\begin{align} \label{FMIT1s final}
		\F[MIT,s][(1)]
			&= \frac{1}{4\pi^2d} \text{Re} \bigg\{ e^{-2i\omega{d}}\bigg[ i\pi G(2d)
				+ \int_{0}^{2d} \frac{\dd{r}}{r}\Big(G(2d+r) e^{-i\omega{r}}
					- G(2d-r) e^{i\omega{r}}\Big) \nonumber \\
			&\qquad \qquad \qquad \qquad \qquad
				+ \int_{2d}^{\infty} \frac{\dd{r}}{r}G(2d+r) e^{-i\omega{r}}
					\bigg] \bigg\},
	\end{align}
where $G$ is again given by~\eqref{G(y)}. 
Combining \eqref{FMIT1r final} and \eqref{FMIT1s final} gives \eqref{Fsub{MIT1}} in the main text. 

This completes the verification of \eqref{FMIT}--\eqref{G(y)} in Section~\ref{Section Static MIT Wall}. 

What remains is to consider the large gap asymptotic form of each of the three terms in~\eqref{FMIT}. 

For $\F[NW]$, the asymptotic form is stated at the end of Section~\ref{Section Absence of Wall}. 

For $\F[MIT][(1)]$, we rewrite \eqref{Fsub{MIT1}} as
	\begin{align} \label{Fsub{MIT1} large energy gap}
		\F[MIT][(1)]
			&= \frac{1}{4\pi^2d} \text{Re} \bigg\{ e^{-2i\omega{d}}\bigg[ 2i\pi G(2d) \Theta(-\omega)
				+ \int_{0}^{\infty} \dd{r} \cos\omega{r} \, 
					\frac{G(2d+r) - G(2d-r)}{r}
						\nonumber \\
			&\qquad \qquad \qquad
				- i\int_{0}^{\infty} \dd{r} \sin\omega{r} \, 
					\frac{G(2d+r) + G(2d-r) - 2G(2d)}{r} \bigg] \bigg\} , 
	\end{align}
by adding and subtracting $2G(2d)\sin(\omega{r})/r$ under the $r$-integral and using the identity $\int_{0}^{\infty} \sin(\omega{x})/x \dd{x} = (\pi/2) \sgn{\omega}$. The method of repeated integration by parts, integrating the trigonometric factor \cite{Wong:2001}, then shows that the integral terms in \eqref{Fsub{MIT1} large energy gap} are $\BigOinf{|\omega|^{-1}}$.

For $\F[MIT][(2)]$, we use the method of repeated integration by parts in~\eqref{Fsub{MIT2}}, 
integrating the trigonometric factor. Using the evenness of~$G$, we find that $\F[MIT][(2)] = \BigOinf{|\omega|^{-1}}$. 

Combining these results gives the asymptotic formula~\eqref{F{MIT} large energy gap limit}.

\subsection{General Evolving Wall} \label{Appendix Evolving Wall}
	
For the general evolving wall, 
the regularised two-point function is given for $z, z' > 0$ by
	\begin{align} \label{General Evolving Wall Two point Function}
		&W_{\epsilon,\text{GEW}}^{(2,\bar{2})}(u,v;u',v') \nonumber \\
			&\quad = \cos\left([\varphi(u)-\varphi(u')]+[\varphi_{1}(u)-\varphi_{1}(u')]\right)
					W_{\epsilon,\text{NW}}^{(2,\bar{2})} \cos\theta(u)\cos\theta(u')
						\nonumber \\
			&\quad \qquad
				+ \cos\left([\varphi(u)-\varphi(u')]+[\varphi_{2}(u)-\varphi_{2}(u')]\right)
					W_{\epsilon,\text{NW}}^{(2,\bar{2})} \sin\theta(u)\sin\theta(u')
						\nonumber \\
			&\quad \qquad
				+ \cos\left([\varphi(u)+\varphi(u')]+[\varphi_{2}(u)+\varphi_{2}(u')]\right)
					W_{\epsilon,\text{MIT}_{1}}^{(2,\bar{2})} \sin\theta(u)\sin\theta(u')
						\nonumber \\
			&\quad \qquad
				+ 2\Re{e^{i[\varphi(u)-\varphi(u')]+i[\varphi_{2}(u)-\varphi_{2}(u')]}
					W_{\epsilon,\text{GEW}_{1}}^{(2,\bar{2})}} \sin\theta(u)\sin\theta(u')\nonumber \\
			&\quad \qquad
				+ 2\Re{e^{-i[\varphi(u)+\varphi(u')]-i[\varphi_{2}(u)+\varphi_{2}(u')]}
					W_{\epsilon,\text{GEW}_{2}}^{(2,\bar{2})}} \sin\theta(u)\sin\theta(u'),
	\end{align}
where
	\begin{align}
		W_{\epsilon,\text{NW}}^{(2,\bar{2})}(u,v;u',v')
			&= - \frac{1}{2\pi^{2}}
				\frac{1}{ ( u - u' - i \epsilon ) ( v - v' - i \epsilon ) }, \\
		W_{\epsilon,\text{MIT}_{1}}^{(2,\bar{2})}(u,v;u',v')
			&= \frac{1}{2\pi^{2}}
				\frac{1}{ ( u - v' - i \epsilon ) ( v - u' - i \epsilon ) }, \\
		W_{\epsilon,\text{GEW}_{1}}^{(2,\bar{2})}(t,z;t',z')
			&= \frac{1}{4\pi^2}
				\frac{1}{4zz' + 2i\epsilon(z-z') + \epsilon^2}, \\
		W_{\epsilon,\text{GEW}_{2}}^{(2,\bar{2})}(t,z;t',z')
			&= \frac{1}{4\pi^2}
				\frac{1}{4zz' + 2i\epsilon(z+z') - \epsilon^2}.
	\end{align}
The first three terms in \eqref{General Evolving Wall Two point Function} come from the first term of~\eqref{Vacuum two-point function in terms of propagators}, and the last two terms in \eqref{General Evolving Wall Two point Function} come from the second term of~\eqref{Vacuum two-point function in terms of propagators}. 
$W_{\epsilon,\text{NW}}^{(2,\bar{2})}$ and $W_{\epsilon,\text{MIT}_{1}}^{(2,\bar{2})}$
are respectively the first and second term in the eternal MIT wall two-point function~\eqref{Static MIT wall two-point function}, 
$W_{\epsilon,\text{NW}}^{(2,\bar{2})}$ being the no-wall two-point function. 

For a detector on the trajectory~\eqref{Detector Trajectory}, 
substituting \eqref{General Evolving Wall Two point Function} into \eqref{Causal response function} gives
	\begin{align} \label{F{GEW}}
		\F[GEW] = \F[GEW][(0)] + \F[GEW][(1)] + \F[GEW][(2)],
	\end{align}
where
\begin{align}
		\F[GEW][(0)]
			&= \sum_{j = 1}^{4} \lim_{\epsilon \to 0_{+}} \frac{-1}{\pi^2}
				\int_{-\infty}^{\infty} \dd{u} \chi_{j}(u) \int_{0}^{\infty} \dd{s} \chi_{j}(u-s)
					\Re{\frac{e^{-i\omega{s}}}{(s-i\epsilon)^{2}}},
						\label{Fsub{GEW}0 Modified} \\
		\F[GEW][(1)]
			&= \lim_{\epsilon \to 0_{+}} \frac{1}{\pi^2} \int_{-\infty}^{\infty} \dd{u} \chi_{3}(u) \int_{0}^{\infty} \dd{s} \chi_{3}(u-s)
					\Re{\frac{e^{-i\omega{s}}}{(s-i\epsilon)^{2}-4d^{2}}}
						\nonumber \\
			&\quad - \lim_{\epsilon \to 0_{+}} \frac{1}{\pi^2} \int_{-\infty}^{\infty} \dd{u} \chi_{4}(u) \int_{0}^{\infty} \dd{s} \chi_{4}(u-s)
					\Re{\frac{e^{-i\omega{s}}}{(s-i\epsilon)^{2}-4d^{2}}},
						\label{Fsub{GEW}1 Modified} \\
\F[GEW][(2)]
			&= \frac{1}{2\pi^2d^2} \int_{-\infty}^{\infty} \dd{u} \chi_{3}(u) \int_{0}^{\infty} \dd{s} \chi_{3}(u-s)
					\cos(\omega{s}),  \label{Fsub{GEW}2 Modified}
\end{align}
and $\chi_{j}$ for $j = 1,2,3,4$ are the modulated switching functions 
\eqref{GEW Modified Switch 1}--\eqref{GEW Modified Switch 4}. 
In~\eqref{Fsub{GEW}2 Modified}, the $\epsilon\to0_+$ limit has already been taken.

Equations 
\eqref{F{GEW}}--\eqref{Fsub{GEW}2 Modified}
differ from 
\eqref{FMIT temporary}--\eqref{eq:app:fmit2} 
only in that the single switching function $\chi$ 
has been replaced by the modulated switching functions~$\chi_j$, 
all of which are smooth with compact support. 
Proceeding as with \eqref{FMIT temporary}--\eqref{eq:app:fmit2}, and using the identity 
$\sum_{j = 1}^{4} [\chi_{j}(\tau)]^2 = [\chi(\tau)]^2$, leads to 
\eqref{Detector Response to GEW}--\eqref{DeltaFsub{GEW}1} in the main text.

\subsection{Wall Creation} \label{Appendix Wall Appearance}

For the wall creation scenario of Section~\ref{Section Wall Appearance and Disappearance}, 
the detector's response $\F[WC]$ is obtained by inserting 
\eqref{Modified Switch 1 for Wall Appearance}--\eqref{Modified Switch 4 for Wall Appearance}
into \eqref{eq:Gjy}--\eqref{Detector Response to GEW Large Energy Gap}. 
We shall here verify that the asymptotic form of $\F[WC]$ as $\delta\to0$, with everything else fixed, is given by 
\eqref{F{RWC}}--\eqref{H_{1}(y)}. 

In the decomposition~\eqref{Detector Response to GEW}, 
consider first the term $\F[GEW][(0)]$ \eqref{Fsub{GEW}0 Modified Regulator-free}, 
whose $\delta$-dependent part 
$\Delta\F[GEW][(0)]$ is given by~\eqref{DeltaFsub{GEW}0}. 
Taking the sum over $j$ under the integrals gives  
\begin{align}
		\sum_{j = 1}^{4} \chi_{j}(u)[\chi_{j}(u)-\chi_{j}(u-s)]
			&= \chi(u)[\chi(u) - \chi(u-s)]
				+ \chi(u)\chi(u-s)\hat{\chi}(u,s),
	\end{align}
where
\begin{align}
\hat{\chi}(u,s) := 1-\cos(2n[\theta(u)-\theta(u-s)])\cos(\theta(u)-\theta(u-s)), 
\label{eq:chihat-def}
\end{align}
and $\theta(t)$ is given by \eqref{Condition of Theta for Wall Appearance}. 
Substituting this in \eqref{Fsub{GEW}0 Modified Regulator-free}, 
and denoting the resulting quantity by $\F[WC][(0)]$, we obtain 
\begin{align} \label{Fsub{WC}0 Rapid Creation Limit}
		\F[WC][(0)]
			&= \F[NW] + \Delta\F[RWC][(0)] , 
\\
\label{Delta-Fsub{WC}0 Rapid Creation Limit}
\Delta\F[RWC][(0)]
& = \frac{1}{\pi^2} \int_{-\infty}^{\infty} \dd{u} \chi(u) \int_{0}^{\infty} \dd{s} \chi(u-s)\cos(\omega{s}) \frac{\hat{\chi}(u,s)}{s^{2}} , 
\end{align}
where the interchange of the integrals in \eqref{Delta-Fsub{WC}0 Rapid Creation Limit} 
is justified by absolute convergence of the double integral, observing that 
${\hat{\chi}(u,s)}/{s^{2}}$ is bounded as $s\to0$. 

Using \eqref{Condition of Theta for Wall Appearance} and \eqref{eq:chihat-def}, 
we see that the contribution to 
$\Delta\F[RWC][(0)]$ \eqref{Delta-Fsub{WC}0 Rapid Creation Limit} from $u \le t_0$ vanishes, 
and 
the contribution from $t_0 \le u \le t_0+\delta$ is~$\BigO{\delta}$. 
In the contribution from $t_0+\delta \le u$, the part from $s\le u - t_0 - \delta$ vanishes, 
the part from $u - t_0 - \delta \le s \le u - t_0$ is~$\BigO{\delta}$, 
and the part from $u - t_0 \le s$ has $\hat{\chi}(u,s) = 1$. This gives 
\begin{align}
		\Delta\F[RWC][(0)]
			&= \frac{1}{\pi^2} \int_{t_{0}+\delta}^{\infty} \dd{u} \chi(u)
					\int_{u-t_{0}}^{\infty} \dd{s} \chi(u-s)\frac{\cos(\omega{s})}{s^2}
						+ \BigO{\delta} 
\notag\\
		&= \Delta\Fconst[RWC,ind][(0)] - \frac{1}{\pi^2} \int_{t_{0}+\delta}^{\infty} \dd{u} \chi(u)
					\int_{u-t_{0}}^{\infty} \dd{s} \chi(u-s)\frac{1-\cos(\omega{s})}{s^2}
				+ \BigO{\delta} 
\notag\\
& = \Delta\Fconst[RWC,ind][(0)] + \F[RWC][(0)] + \BigO{\delta}, 
\label{DeltaFRWC0}
\end{align}
where 
	\begin{align}
\Delta\Fconst[RWC,ind][(0)]
			&=	\frac{1}{\pi^2} \int_{t_{0}+\delta}^{\infty} \dd{u} \chi(u)
					\int_{u-t_{0}}^{\infty} \dd{s} \frac{\chi(u-s)}{s^2},
						\label{DeltaFsub{RWC,ind}} \\
		\F[RWC][(0)]
			&= -\frac{1}{\pi^2} \int_{0}^{\infty} \dd{s} \frac{1-\cos(\omega{s})}{s^2} f(s),
					\label{Fsub{RWC}0app}
	\end{align}
and $f(s)$ is given in~\eqref{f(s)}. 
In~\eqref{DeltaFRWC0}, the second equality arises by adding and subtracting $\chi(u-s)/s^{2}$ under the $s$-integral; 
the third equality arises by observing that in the $\omega$-dependent term the lower limit of the $u$-integral can be replaced by $t_0$ at the expense of an $\BigO{\delta}$ error, interchanging the integrals, and changing the inner integration variable to give the manifestly odd expression \eqref{f(s)} for~$f(s)$. 
The subscript ``ind'' in $\Delta\Fconst[RWC,ind][(0)]$ 
refers to independence of~$\omega$. 

$\F[RWC][(0)]$ \eqref{Fsub{RWC}0app} does not depend on~$\delta$. 
To isolate the $\delta$-dependence of $\Delta\Fconst[RWC,ind][(0)]$ \eqref{DeltaFsub{RWC,ind}}, 
we integrate the inner integral in \eqref{DeltaFsub{RWC,ind}} by parts, obtaining 
\begin{align} \label{DeltaFsub{RWC,ind} OTW}
		\Delta\Fconst[RWC,ind][(0)]
			&=\frac{\chi(t_{0})}{\pi^2} \int_{t_{0}+\delta}^{\infty} \dd{u} \frac{\chi(u)}{u-t_{0}}
				- \frac{1}{\pi^2} \int_{t_{0}}^{\infty} \dd{u} \chi(u)
					\int_{u-t_{0}}^{\infty} \dd{s} \frac{\chi'(u-s)}{s}
						\nonumber \\
			&\qquad +
				\frac{1}{\pi^2} \int_{t_{0}}^{t_{0}+\delta} \dd{u} \chi(u)
					\int_{u-t_{0}}^{\infty} \dd{s} \frac{\chi'(u-s)}{s}.
\end{align}
In \eqref{DeltaFsub{RWC,ind} OTW}, the last term is 
$\BigO{\delta \ln\delta}$, and the second term is independent of $\delta$. 
Integrating the first term by parts gives 
	\begin{align} \label{DeltaFsub{RWC,ind} Final}
		\Delta\Fconst[RWC,ind][(0)]
			&= - \frac{[\chi(t_{0})]^2}{\pi^2} \ln(\mu\delta)
				- \frac{\chi(t_{0})}{\pi^2}\int_{t_{0}}^{\infty} \dd{u} \chi'(u)\ln\bigl(\mu(u-t_{0})\bigr) 
					\nonumber \\
			&\qquad
				- \frac{1}{\pi^2} \int_{t_{0}}^{\infty} \dd{u} \chi(u)
					\int_{u-t_{0}}^{\infty} \dd{s} \frac{\chi'(u-s)}{s}
				+ \BigO{\delta\ln\delta}, 
	\end{align}
where in the first term we have replaced $\chi(t_{0}+\delta)$ by $\chi(t_{0})$ 
and in the second term we have replaced the lower integration limit $t_0+\delta$ by~$t_0$, 
each at the expense of an $\BigO{\delta\ln\delta}$ error. 
In the first two terms in \eqref{DeltaFsub{RWC,ind} Final} 
we have also introduced a positive constant~$\mu$, of dimension inverse length, 
for dimensional consistency; the value of $\mu$ cancels out from the sum of these two terms. 

Combining these observations gives the first four terms in $\F[WC]$~\eqref{F{RWC}}. 

Consider next the term $\F[GEW][(1)]$ \eqref{Fsub{GEW}1 Modified Regulator-free} with 
$\Delta\F[GEW][(1)]$ \eqref{DeltaFsub{GEW}1}, for the modified switching functions \eqref{Modified Switch 1 for Wall Appearance}--\eqref{Modified Switch 4 for Wall Appearance}; 
we denote these terms respectively by $\F[WC][(1)]$ and~$\Delta\F[WC][(1)]$. 
As we are considering the limit $\delta \to 0$, we may assume $\delta < 2d$. 
It follows from \eqref{Modified Switch 4 for Wall Appearance} that $G_{4}(z) = 0$ for $\delta \le |z|$; 
hence, $G_4(2d)$ and $G_4(2d+r)$ do not contribute in 
\eqref{Fsub{GEW}1 Modified Regulator-free} and~\eqref{DeltaFsub{GEW}1}, 
while $G_4(2d-r)$ in \eqref{DeltaFsub{GEW}1} 
contributes only for $2d-\delta < r <  2d + \delta$, and hence by~$\BigO{\delta}$.
For $G_3$, we note from \eqref{Modified Switch 3 for Wall Appearance} that 
$G_{3}(2d) = H_{1}(2d) + \BigO{\delta}$, where $H_{1}(y)$ is given by \eqref{H_{1}(y)}.
Using similar considerations for all the contributions from $G_3$ in \eqref{DeltaFsub{GEW}1} 
gives \eqref{Fsub{RWC}1} plus errors of order $\BigO{\delta}$. 
Collecting the results, we get $\F[WC][(1)] = \F[RWC][(1)] + \BigO{\delta}$, 
where $\F[RWC][(1)]$ is as given by \eqref{Fsub{RWC}1}.

Consider finally the term 
$\F[GEW][(2)]$ \eqref{Fsub{GEW}2 Modified Regulator-free} with $\chi_3$ given by \eqref{Modified Switch 3 for Wall Appearance}. Similar techniques lead to $\F[RWC][(2)]$ given in \eqref{Fsub{RWC}2} plus an $\BigO{\delta}$ error. 

Combining all these observations gives $\F[WC]$~\eqref{F{RWC}}. 
	
	\addcontentsline{toc}{section}{References}
	\bibliographystyle{JHEP}
	\bibliography{Articles,Books,InCollection,Online,PhDThesis}

\end{document}